\documentclass[aps,prb,twocolumn,showpacs,citeautoscript,reprint,longbibliography,superscriptaddress]{revtex4-2}
\usepackage{graphicx}
\usepackage{bm}
\usepackage{natbib}
\usepackage{amsmath}
\usepackage{xfrac}
\usepackage{nicefrac}
\usepackage[colorlinks=true, urlcolor=blue, linkcolor=blue, citecolor=blue, pdfborder={0 0 0}]{hyperref}
\usepackage[caption=false]{subfig}
\usepackage{cleveref}
\crefname{figure}{Fig.}{Figs}
\crefname{table}{Table}{Tables}
\crefname{section}{Sec.}{Sections}
\usepackage{dcolumn}
\usepackage{ulem}

\begin{document}
\preprint{Version: v4d}

\title{Spin transport at finite temperatures: A first-principles study for ferromagnetic$|$nonmagnetic interfaces}

\author{Kriti Gupta}
\affiliation{Faculty of Science and Technology and MESA$^+$ Institute for Nanotechnology, University of Twente, P.O. Box 217,
		7500 AE Enschede, The Netherlands}
\author{Rien J. H. Wesselink}
\affiliation{Faculty of Science and Technology and MESA$^+$ Institute for Nanotechnology, University of Twente, P.O. Box 217,
		7500 AE Enschede, The Netherlands}
\author{Zhe Yuan}
\email[Email: ]{zyuan@bnu.edu.cn}
\affiliation{The Center for Advanced Quantum Studies and Department of Physics, Beijing Normal University, 100875 Beijing, China}
\author{Paul J. Kelly\thanks{corresponding author}}
\email[Email: ]{P.J.Kelly@utwente.nl}
\affiliation{Faculty of Science and Technology and MESA$^+$ Institute for Nanotechnology, University of Twente, P.O. Box 217,
		7500 AE Enschede, The Netherlands}
\affiliation{The Center for Advanced Quantum Studies and Department of Physics, Beijing Normal University, 100875 Beijing, China}

\date{\today}

\begin{abstract}
Symmetry lowering at an interface leads to an enhancement of the effect of spin-orbit coupling and to a discontinuity of spin currents passing through the interface. This discontinuity is characterized by a ``spin-memory loss'' (SML) parameter $\delta$ that has only been determined directly at low temperatures. Although $\delta$ is believed to be significant in experiments involving interfaces between ferromagnetic and nonmagnetic metals, especially heavy metals like Pt, it is more often than not neglected to avoid introducing too many unknown interface parameters in addition to often poorly known bulk parameters like the spin-flip diffusion length $l_{\rm sf}$. In this work, we calculate $\delta$ along with the interface resistance $AR_{\rm I}$ and the spin-asymmetry parameter $\gamma$ as a function of temperature for Co$|$Pt and Py$|$Pt interfaces where Py is the ferromagnetic Ni$_{80}$Fe$_{20}$ alloy, permalloy. We use first-principles scattering theory to calculate the conductance as well as local charge and spin currents, modeling temperature-induced disorder with frozen thermal lattice and, for ferromagnetic materials, spin disorder within the adiabatic approximation. The bulk and interface parameters are extracted from the spin currents using a Valet-Fert model generalized to include SML.
\end{abstract}


\maketitle


\section{Introduction}
\label{Sec:Intro}

Experiments in the field of spintronics are almost universally interpreted using semiclassical transport theories \cite{Gijs:ap97, Barthelemy:99, Brataas:prp06, Bass:jmmm16}. In such phenomenological theories, based upon the Boltzmann \cite{Valet:prb93} or diffusion equations \cite{Johnson:prb87, vanSon:prl87, *vanSon:prl88}, the material and structure dependence enters via a multitude of parameters. 
For example, the transport properties of a bulk material are characterized in terms of a resistivity $\rho$, a polarization (or spin-asymmetry) parameter $\beta$ that vanishes for nonmagnetic materials, and a spin-flip diffusion length (SDL) $l_{\rm sf}$. A NM$|$FM interface between nonmagnetic (NM) and ferromagnetic (FM) metals is characterized analogously in terms of an interface resistance $AR_{\rm I}$, a polarization $\gamma$ and a spin memory loss (SML) parameter $\delta$. To describe the transport in noncollinearly aligned ${\rm FM|NM|FM'}$ spin valves it is necessary to introduce an additional parameter, a complex, so-called spin-mixing conductance $G_{\uparrow\downarrow}$ \cite{Brataas:prl00, Brataas:prp06, Weiler:prl13}. 
In ferromagnetic materials the spin-orbit coupling (SOC) and conductivity polarization lead to a Hall effect in the absence of an external magnetic field that is characterized by the ``anomalous Hall angle'' $\Theta_{\rm aH}$. 
In (heavy) nonmagnetic elements the SOC gives rise to the spin Hall effect (SHE) \cite{Dyakonov:zetf71, *Dyakonov:pla71, Hirsch:prl99, Zhang:prl00, Hoffmann:ieeem13, Sinova:rmp15} whereby an electric current leads to the generation of a transverse spin current. The SHE is characterized in terms of the spin Hall angle (SHA) $\Theta_{\rm sH}$ that is the ratio of the spin current (measured in units of $\hbar/2$) to the charge current (measured in units of the electron charge $-|e|$). For interfaces an interface SHA $\Theta_{\rm sH}^{\rm I}$ can be defined by analogy \cite{WangL:prl16}. 
Phenomenological theories ultimately aim to relate currents of charge ${\bf j}_c$ and spin ${\bf j}_{s\alpha}$ to gradients of the chemical potential $\mu_c$ and spin accumulation ${\bm \mu}_{s\alpha}$ in terms of the above parameters but tell us nothing about the values of the parameters for particular materials or combinations of materials \cite{Bass:jmmm16}. Here $\alpha$ labels the spin component. 

It has turned out to be remarkably difficult to measure many of the parameters described above quantitatively \cite{Bass:jpcm07, Bass:jmmm16}, especially at other than low temperatures. In particular, virtually nothing is known about the interface parameters $AR_{\rm I}$, $\gamma$ and $\delta$ at room temperature because, unless the sample cross sections are reduced by structuring \cite{Gijs:prl93}, the interface resistance is swamped by other resistances. The use of superconducting leads restricts studies to the low superconducting critical temperatures of commonly used metals like Al or Nb \cite{Pratt:prl91}. At these low temperatures transport properties are strongly extrinsic but little is known about the nature of the bulk disorder that gives rise to the observed diffuse transport. The situation with respect to interface disorder is even worse because so little is known about it on an atomic level.

Ten years ago only a handful of measurements had been made of the SDL \cite{Bass:jpcm07} or of the SHA \cite{Hoffmann:ieeem13, Sinova:rmp15}. The advent of nonlocal spin injection \cite{Valenzuela:nat06, Kimura:prl07, Vila:prl07} and spin-pumping (SP) \cite{Saitoh:apl06, Mosendz:prl10, Mosendz:prb10, Azevedo:prb11} techniques has allowed the SHA to be determined by means of the inverse SHE (ISHE). Alternatively, spin currents generated by the SHE can be used to drive the precession of a magnetization by the spin-transfer torque (STT) that is monitored using ferromagnetic resonance (FMR) \cite{Ando:prl08, Liu:prl11}. These innovations have changed the situation radically over the past ten years yielding a host of very disparate room temperature results for $l_{\rm sf}$ and $\Theta_{\rm sH}$ \cite{Hoffmann:ieeem13, Sinova:rmp15}. The new measurement techniques make use of interfaces through which spin must flow in order to be detected. Though attempts have been made to take the interface effects described above into account \cite{Rojas-Sanchez:prl14, ZhangW:natp15, Nguyen:prl16, Tao:sca18, Berger:prb18b}, this has yet to be done systematically by determining all parameters for consistent sets of samples. If anything, it has led to an increase in the spread of values reported for the key parameters $l_{\rm sf}$ and $\Theta_{\rm sH}$ for e.g. Pt \cite{Wesselink:prb19}.  
   
To accurately estimate the generation and detection efficiency of spin currents, which is one of the key concerns of spintronics, it is necessary to carefully characterize the samples used to measure all of the material parameters described above. 
The ultimate goal is to be able to make efficient spintronics devices at finite temperatures \cite{Manipatruni:nat18}, where intrinsic scattering mechanisms play an important role. The earliest attempt \cite{Rojas-Sanchez:prl14} to include spin memory loss in determining  $l_{\rm sf}$ and $\Theta_{\rm sH}$ for Pt at room temperature relied on estimates available at 4.2 K from magnetoresistance experiments \cite{Nguyen:jmmm14}. The almost complete lack of information about how interface parameters might depend on temperature motivated the work that is presented here.

To quantitatively describe the magnetic and transport properties of transition metals requires taking into account their degenerate electronic structures. For the layered structures that form the backbone of spintronics, the most promising way to combine complex electronic structures with transport theory is to use scattering theory \cite{Datta:95} formulated either in terms of nonequilibrium Green's functions or wave-function matching \cite{Brataas:prp06} that are equivalent in the linear response regime \cite{Khomyakov:prb05}. With few exceptions \cite{Dolui:prb17}, the attempts that have been made to address interface properties have been based upon circuit theory whereby quantum mechanical transmission matrices form boundary conditions to match solutions of semiclassical Boltzmann or diffusion equations on either side of the interface. Such calculations have been used to calculate interface resistances \cite{Schep:prb97, Xia:prb01, Bauer:jpd02, Stiles:prb00, Zwierzycki:prb03, Xu:prl06}, mixing conductances \cite{Xia:prb02, Zwierzycki:prb05, Turek:jpcm07}, the spin-dependent transparency of FM$|$superconducting interfaces \cite{Xia:prl02} and recently the SML \cite{Belashchenko:prl16}. In all of these applications, it is tacitly assumed that the interface properties are temperature independent. A priori it is not clear what the effects of temperature will be. It was found that the interface resistance could be increased (Co$|$Cu) or reduced (Fe$|$Cr) by interface disorder depending on the Fermi surfaces on either side of the interface \cite{Xia:prb01, Bauer:jpd02}.

We recently demonstrated a simple and effective way of including temperature-induced lattice and spin disorder in the adiabatic approximation \cite{LiuY:prb15, Starikov:prb18} in first-principles scattering calculations. Using the Landauer-B{\"{u}}ttiker formalism, the resistivity can be extracted from calculations of the conductance as a function of the length $L$ of the scattering region \cite{Starikov:prl10, Starikov:prb18}. By calculating local spin currents \cite{Wesselink:prb19} (and chemical potentials) from the scattering theory results, we can make direct contact with experiments interpreted using the Valet-Fert (VF) formalism that is expressed in terms of these same variables \cite{Valet:prb93}. By focussing on local currents we showed how the interface effects that are always present in scattering calculations could be factored out, and illustrated the approach with calculations of the room temperature spin-flip diffusion lengths of Pt and Py, the polarization of Py and the SHA for Pt \cite{Wesselink:prb19}. 

In this paper, we extend the above approach to study the temperature dependence of the  transport properties of Co$|$Pt and Py$|$Pt interfaces. Because the interface and bulk parameters are inextricably coupled, we will first determine the bulk parameters for Pt ($\rho$, $l_{\rm sf}$), Co and Py ($\rho$, $\beta$ and $l_{\rm sf}$) before determining the three parameters used to characterize collinear spin transport through an interface ($AR_{\rm I}$, $\gamma$ and $\delta$). We will be able to address how the thermal disorder and magnetic ordering of Py versus Co influence these parameters. We also investigate how proximity-induced magnetization in Pt influences the interface.
A short report of this work appeared in Ref.~\cite{Gupta:prl20}.

The plan of the paper is as follows. 
In \cref{sec:Methods} we summarize how the VF model is extended to include the effect of SOC at interfaces (\cref{sec:VF}) and describe how it will be used to extract interface parameters (\cref{sec:VFE}). 
In \cref{sec:Calculations} we briefly summarize the first-principles scattering theory \cite{Xia:prb06, Starikov:prb18, Wesselink:prb19} and give details of how fully relaxed Co$|$Pt and Py$|$Pt interface geometries are constructed, how temperature is incorporated in the adiabatic approximation, how the necessary atomic sphere potentials are calculated and how the length of the scattering system is determined. 
In \cref{sec:Results}, we determine the temperature dependence of the Pt, Co and Py bulk transport parameters (\cref{sec:BM}) and of the interface transport parameters for Co$|$Pt and Py$|$Pt interfaces (\cref{sec:Ints}). To the best of our knowledge, this is the first attempt to evaluate the effect of thermal disorder on interfaces. \cref{Sec:Exp} contains a discussion of the results presented in the previous section. 

\section{Methods}
\label{sec:Methods}

The original VF model parametrized interfaces in terms of the spin-dependent interface resistances $R_{\uparrow}$ and  $R_{\downarrow}$. The model was extended by Fert and Lee \cite{Fert:prb96b} to include interface SOC in the form of a spin-flip interface resistance. It was reformulated in terms of the SML parameter $\delta$ by Baxter {\it et al.} for NM$|$NM$'$ interfaces between two nonmagnetic metals \cite{Baxter:jap99}, and by Eid {\it et al.} for NM$|$FM interfaces between nonmagnetic and ferromagnetic metals \cite{Eid:prb02}. We summarize this generalized VF model in the next subsection and then extract the boundary conditions for a geometrically sharp NM$|$FM interface.

\subsection{Valet-Fert model}
\label{sec:VF}

Starting from the Boltzmann formalism, Valet and Fert derived the following equations to describe a spin current flowing along the $z$ direction perpendicular to the interface plane (CPP) for an axially symmetric geometry
\begin{subequations}
\begin{align}
\frac{\partial^2 \mu_s}{\partial z^2}&=\frac{\mu_s}{l_{\rm sf}^2} 
\label{eq:diffusion}\\
j_{\sigma}(z)&=-\frac{1}{e \rho_{\sigma}}\frac{\partial\mu_{\sigma}}{\partial z} 
\label{eq:ohm}
\end{align}
\end{subequations}
where $\sigma= \, \uparrow$ and $\downarrow$ for majority and minority spins, respectively, $\mu_s = \mu_\uparrow - \mu_\downarrow$ is the spin accumulation and $j_s = j_\uparrow - j_\downarrow$ is a spin current density. Equations \eqref{eq:diffusion} and \eqref{eq:ohm} can be solved for $\mu_\uparrow$, $\mu_\downarrow$, $j_\uparrow$ and $j_\downarrow$ making use of the condition that the total current density $j=j_\uparrow+j_\downarrow$ is conserved in one-dimensional transport. The solutions are 
\begin{subequations}
\begin{align}
\mu_s(z)&=A e^{z/l_{\rm sf}}+B e^{-z/l_{\rm sf}} \\
\widehat{j_s}(z) &= \beta - \frac{(1-\beta^2)}{2ej \rho l_{\rm sf}}
                 \Big[A e^{z/l_{\rm sf}}-B e^{-z/l_{\rm sf}}\Big]
\end{align}
\label{eq:VFsol}
\end{subequations}
where $\widehat{j_s}(z) \equiv j_s/j $ is the normalized spin-current density and $\beta$ is the spin-asymmetry (or polarization) parameter
\begin{equation}
\beta = \frac{\rho_{\downarrow} - \rho_{\uparrow}}
             {\rho_{\downarrow} + \rho_{\uparrow}}  .
\end{equation}
Instead of using the resistivity $\rho$ and polarization $\beta$, the spin-dependent resistivities $\rho_{\downarrow}$ and $\rho_{\uparrow}$ are frequently used with 
$\rho_{\uparrow} + \rho_{\downarrow} =4 \rho^*$ and 
$\rho_{\uparrow} - \rho_{\downarrow} =4 \rho^* \beta$ where the auxiliary quantity $\rho^* = \rho/(1-\beta^2)$.

\begin{figure}[t]
\includegraphics[width=8.8 cm]{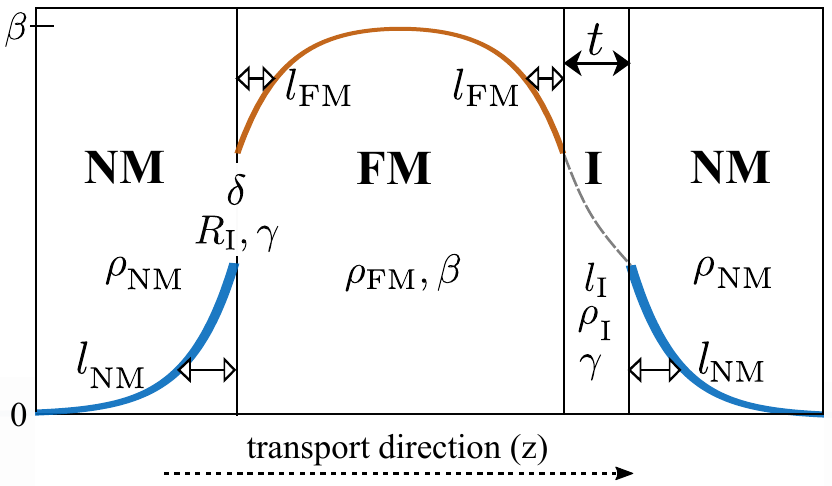}
\caption{Normalized spin current $\widehat{j_s}(z)$ across a ${\rm NM|FM|NM}$ trilayer as described by the VF equations. $\widehat{j_s}(z)$ is assumed to be continuous at the FM$|$I and ${\rm I|NM}$ interfaces between the FM and NM bulk layers and the fictitious bulklike interface (I) layer with thickness $t$. In the limit that $t \rightarrow 0$, a discontinuity occurs in $\widehat{j_s}(z)$ at the left-hand NM$|$FM interface. 
}
\label{figA}
\end{figure}

The coefficients $A$ and $B$ in \eqref{eq:VFsol} are chosen to satisfy appropriate boundary conditions. This is illustrated in \cref{figA} with a sketch of $\widehat{j_s}(z)$ that arises in a diffusive symmetric ${\rm NM|FM|NM}$ trilayer  when a current of electrons is passed from left to right. Far away (measured in units of $l_{\rm NM} \equiv l_{\rm sf}^{\rm NM}$) on the left, the spin current is unpolarized because of the symmetry of the two spin channels in a NM conductor; the solution \eqref{eq:VFsol} therefore only contains an exponentially increasing term. For a sufficiently thick FM (measured in units of $l_{\rm FM} \equiv l_{\rm sf}^{\rm FM}$) in which there is an asymmetry between the up-spin and down-spin channels, $\widehat{j_s}$ saturates to the value $\beta$. Far to the right in the NM material, it becomes zero again and \eqref{eq:VFsol} only contains an exponentially decreasing term. In the central FM material both terms are present. 
The spin and charge currents in the FM and NM bulk layers are characterised in terms of the appropriate resistivities $\rho$ and spin-flip diffusion lengths $l_{\rm sf}$. The asymmetry between the two spin channels in the FM is additionally characterised by the bulk spin asymmetry parameter $\beta$. 

An interface is modelled by introducing a fictitious interface (I) layer with interface resistivity $\rho_{\rm I}$, polarization $\beta_{\rm I} \equiv \gamma$, and SDL $l_{\rm I} \equiv l_{\rm sf}^{\rm I}$. This is illustrated in \cref{figA} for the right-hand ${\rm FM|NM}$ interface  which is shown exploded as an ${\rm FM|I|NM}$ trilayer with a ``bulk-like'' I layer with finite thickness $t$. In this exploded representation, the spin current density is continuous at the FM$|$I and ${\rm I|NM}$ interfaces. For an interface area $A$, an interface resistance is defined as $AR_{\rm I}=\rho_{\rm I} t$ and the SML as $\delta = t / l_{\rm I} $. When $t \rightarrow 0$, a spin current discontinuity occurs at the interface as sketched for the left NM$|$FM interface in \cref{figA}. This discontinuity is attributed to interface spin-flip scattering and described in terms of $\delta$. 
Instead of $AR_{\rm I}$ and $\gamma$ we can use the spin-dependent interface resistances $AR_{\downarrow}$ and $AR_{\uparrow}$ with $AR_{\uparrow} + AR_{\downarrow} =4 AR^*_{\rm I}$, $AR_{\uparrow} - AR_{\downarrow} =4 AR^*_{\rm I} \gamma$ and $AR^*_{\rm I} = AR_{\rm I}/(1-\gamma^2)$.
 
\begin{figure}[t]
	\begin{center}
		\includegraphics[width=8.8 cm]{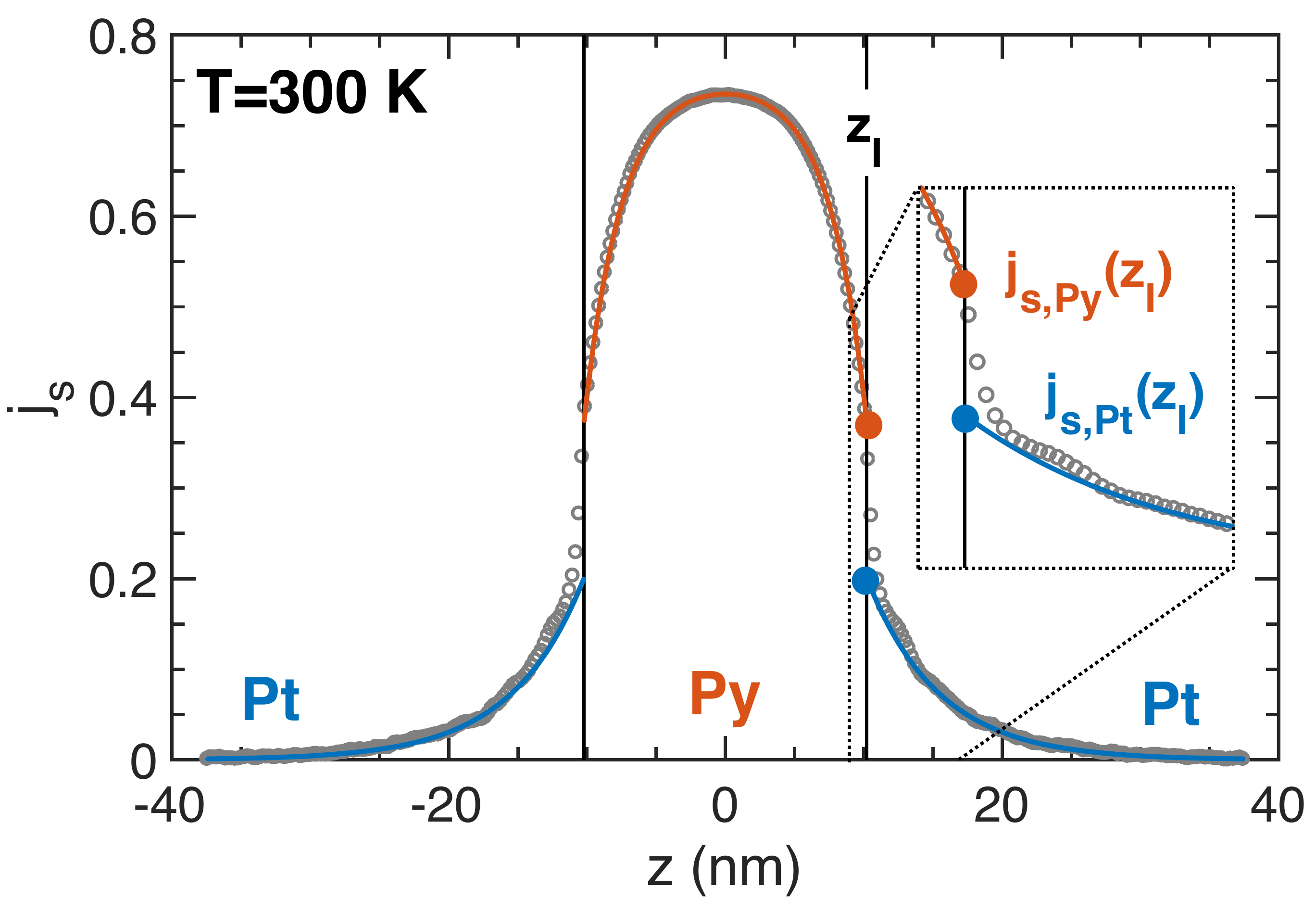}
		\includegraphics[width=8.8 cm]{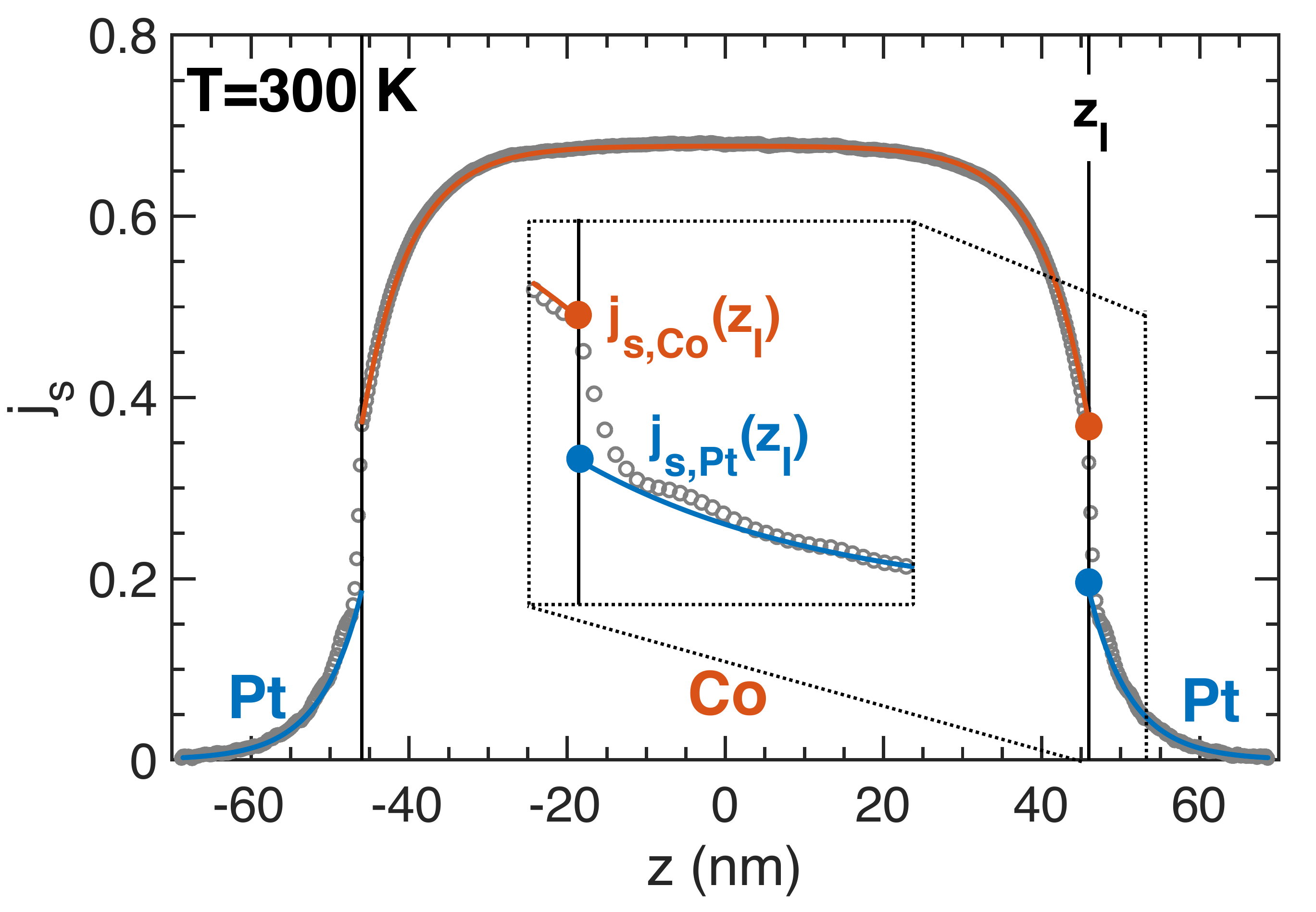}
	\end{center}
\caption{Spin current $\widehat{j}_s(z)$ calculated for (top) Pt$|$Py$|$Pt and (bottom) Pt$|$Co$|$Pt trilayers at 300 K from the results of quantum mechanical scattering calculations. The data is fitted to the VF equations in bulk Pt (blue curves) and Py/Co (orange curve) and extrapolated to the interface $z_{\rm I}$ to obtain the values $\widehat{j}_{s,{\rm Pt}}(z_{\rm I})$ separately for both cases and $\widehat{j}_{s,{\rm Py}}(z_{\rm I})$ and $\widehat{j}_{s,{\rm Co}}(z_{\rm I})$, respectively, which are used to calculate $\delta$ and $\gamma$.}
\label{figB}
\end{figure}

As sketched in \cref{figA}, interface spin-flipping is thus expected to lead to a discontinuity in the spin current. The result of calculating $\widehat{j_s}(z)$ from the output of quantum mechanical scattering calculations \cite{Wesselink:prb19} for (111) oriented diffusive Pt$|$Py$|$Pt and Pt$|$Co$|$Pt trilayers sandwiched between ballistic Cu leads is shown in \cref{figB} where each data point corresponds to a layer of atoms.
The interface parameters cannot be determined simply from the calculated spin current by fitting because $\widehat{j_s}(z)$ depends not only on the three interface parameters but also on five bulk parameters: two parameters for bulk NM and three for bulk FM so $\widehat{j_s}(z)\equiv \widehat{j_s}(\rho_{\rm NM},AR_{\rm I},\rho_{\rm FM},\beta_{\rm I},\beta_{\rm FM},l_{\rm NM},\delta,l_{\rm FM},z)$. Instead, we will first determine the bulk parameters with separate calculations for NM and FM materials. Then, using these parameters we will extrapolate $\widehat{j_s}(z)$ for the NM$|$FM$|$NM trilayer to the interface at $z=z_{\rm I}$ from the NM side to yield $\widehat{j}_{s,{\rm NM}}(z_{\rm I})$ and then from the FM side to yield $\widehat{j}_{s,{\rm FM}}(z_{\rm I})$. This will leave us to determine three unknown interface parameters from two values of $\widehat{j_s}(z_{\rm I})$. $AR_{\rm I}$ can be determined independently by calculating the resistance of a Pt$|$FM$|$Pt trilayer as a function of the thickness of the FM layer leaving us to determine $\delta$ and $\gamma$ from the discontinuity of $\widehat{j_s}(z)$ at the interface. In the following subsection, we will explain how this will be done without having to determine $\mu_s(z)$ explicitly.

\subsection{Interface discontinuity}
\label{sec:VFE}

Viewing an interface as a fictitious bulk-like ``interface'' material transforms a FM$|$NM bilayer of two materials into a FM$|$I$|$NM trilayer. 
The solutions \eqref{eq:VFsol} of the VF equations for the three distinct layers labelled $i=$ FM, I, NM are 
\begin{subequations}
\begin{eqnarray}
\mu_{si}(z)&=&A_ie^{z/l_i}+B_i e^{-z/l_i}\\
j_{si}(z)&=&\beta_i-\frac{(1-\beta^2_i)}{2ej\rho_i l_i}\Big[A_i e^{z/l_i}-B_i e^{-z/l_i}\Big]
\label{eq:js}
\end{eqnarray}
\end{subequations}
and longitudinal spin transport is characterized by the nine transport parameters $\rho_i,~\beta_i,~l_i$ (with $\beta_{\rm NM} \equiv 0$ and $\beta_{\rm I} \equiv \gamma $). The $\widehat{\phantom{.}}$ over the normalized current will be omitted when it does not lead to any confusion.

We want to switch from an FM$|$I$|$NM picture where $\mu_s$ and $j_s$ are continuous everywhere to an FM$|$NM description with discontinuities in $\mu_s(z_{\rm I})$ and $j_s(z_{\rm I})$ at the sharp interface $z=z_{\rm I}$ (\cref{figA}). Continuity at the FM$|$I interface leads to  
\begin{subequations}
\begin{align}
\mu_{s,{\rm FM}}(z_{\rm I})&=A_{\rm I} e^{z_{\rm I}/l_{\rm I}}+B_{\rm I} e^{-z_{\rm I}/l_{\rm I}}\\
j_{s,{\rm FM}}(z_{\rm I})&=\gamma-\frac{(1-\gamma^2)}{2ej\rho_{\rm I} l_{\rm I}}\Big[A_{\rm I} e^{z_{\rm I}/l_{\rm I}}-B_{\rm I} e^{-z_{\rm I}/l_{\rm I}}\Big]
\end{align}
\end{subequations}
and at the I$|$NM interface to 
%
%
\begin{subequations}
\begin{align}
\mu_{s,{\rm NM}}(z_{\rm I}+t)&=A_{\rm I} e^{(z_{\rm I}+t)/l_{\rm I}}+B_{\rm I} e^{-(z_{\rm I}+t)/l_{\rm I}}\\
j_{s,{\rm NM}}(z_{\rm I}+t)&= \gamma- \frac{(1-\gamma^2)}{2ej\rho_{\rm I} l_{\rm I}} \nonumber \\
             &\times \Big[A_{\rm I} e^{(z_{\rm I}+t)/l_{\rm I}}-B_{\rm I} e^{-(z_{\rm I}+t)/l_{\rm I}}\Big].
\end{align}
\end{subequations}
The coefficients $A_{\rm I}$ and $B_{\rm I}$ can be expressed in terms of $\mu_{s,{\rm FM}}(z_{\rm I})$ and $\mu_{s,{\rm NM}}(z_{\rm I}+t)$. Taking the limit $t \rightarrow 0$ results in the expected discontinuity in $\mu_s$ and $j_s$ at the FM$|$NM interface. Substituting $t/l_{\rm I}=\delta$ and $\rho_{\rm I}=AR_{\rm I}/t$ yields
\begin{subequations}
\label{eq:del-gam}
\begin{eqnarray}
j_{s,{\rm FM}}(z_{\rm I}) =\gamma - &&\frac{(1-\gamma^2)\delta}{2ejAR_{\rm I} \sinh \delta}   \nonumber \\              
 \times && \Big[\mu_{s,{\rm NM}}(z_{\rm I}) - \mu_{s,{\rm FM}}(z_{\rm I}) \cosh \delta \Big]   
\end{eqnarray}
\begin{eqnarray}
j_{s,{\rm NM}}(z_{\rm I})=\gamma -&&\frac{(1-\gamma^2)\delta}{2ejAR_{\rm I} \sinh \delta} \nonumber \\  
 \times &&\Big[\mu_{s,{\rm NM}}(z_{\rm I})\cosh \delta - \mu_{s,{\rm FM}}(z_{\rm I})\Big]
\end{eqnarray}
\end{subequations}
%
which is the desired result. In the next paragraph, we specialize to a symmetric NM$|$FM$|$NM trilayer and describe how we will extract the spin-flipping parameter $\delta$ and the spin-asymmetry parameter $\gamma$ for a ${\rm FM}|{\rm NM}$ interface.  

\subsubsection{Symmetric trilayer}

Although we are interested in the properties of a single interface between thermally disordered FM and NM, embedding an FM$|$NM bilayer between ballistic NM$'$ leads would result in an NM$'|$FM$|$NM$|$NM$'$ scattering geometry and a new NM$'|$FM interface with additional interface parameters. Instead, we consider a thermally disordered NM$|$FM$|$NM scattering region embedded between left and right ballistic (NM$'$) leads. The advantages of this geometry are two-fold. (i) The inversion symmetry of the system makes $A_{\rm FM}=-B_{\rm FM}$. (ii) For sufficiently thick NM (and unpolarised leads), the spin currents far from the FM$|$NM interfaces decay to zero allowing us to assume $B_{\rm NM}({\rm left})=A_{\rm NM}({\rm right})=0$. By choosing $z=0$ in the middle of the central FM layer, the expressions for  $j_s(z)$ and $\mu_s(z)$ simplify to
\begin{subequations}
\begin{align}
\mu_{s,{\rm FM}}(z)&=A_{\rm FM} \Big[ e^{z/l_{\rm FM}}- e^{-z/l_{\rm FM}}\Big] \\
j_{s,{\rm FM}}(z)&=\beta-\frac{(1-\beta^2)}{2ej\rho_{\rm FM} l_{\rm FM}}A_{\rm FM} \Big[ e^{z/l_{\rm FM}} + e^{-z/l_{\rm FM}}\Big].
\label{eq:jFM}
\end{align}
\end{subequations}
Combining these to eliminate $A_{\rm FM}$ yields 
\begin{equation}
\mu_{s,{\rm FM}}(z)=2ej \frac{\rho_{\rm FM}l_{\rm FM}} {1-\beta^2} \tanh\bigg(\frac{z}{l_{\rm FM}}\bigg)  \Big[ \beta-j_{s,{\rm FM}}(z) \Big].
\end{equation}
In an analogous manner, the expressions for  $j_s(z)$ and $\mu_s(z)$ in the right NM layer become
\begin{subequations}
\begin{align}
\mu_{s,{\rm NM}}(z)&=B_{\rm NM}e^{-z/l_{\rm NM}} \\
j_{s,{\rm NM}}(z)&=\frac{B_{\rm NM}}{2ej\rho_{\rm NM} l_{\rm NM}} e^{-z/l_{\rm NM}}\label{eq:jNM}  
\end{align}
\end{subequations}
which when combined result in 
\begin{equation}
\mu_{s,{\rm NM}}(z)=2ej\rho_{\rm NM} l_{\rm NM}j_{s,{\rm NM}}(z).
\end{equation}

We can now replace $\mu_{s,{\rm FM}}(z_{\rm I})$ and $\mu_{s,{\rm NM}}(z_{\rm I})$ in equations \eqref{eq:del-gam} with $j_{s,{\rm FM}}(z_{\rm I})$ and $j_{s,{\rm NM}}(z_{\rm I})$ . The choice of origin at the center of the FM layer means that $z_{\rm I}=L_{\rm FM}/2$. Our final expressions for the values of the spin current at a FM$|$NM interface in terms of the transport parameters are
\begin{widetext}
\begin{subequations}
\label{eq:jsi}
\begin{align}
j_{s,{\rm FM}}(z_{\rm I})&=\gamma-\frac{(1-\gamma^2)\delta}{AR_{\rm I} \sinh \delta}
    \bigg[\rho_{\rm NM}l_{\rm NM}j_{s,{\rm NM}}(z_{\rm I}) 
             -\cosh \delta \, \frac{\rho_{\rm FM}l_{\rm FM}} {1-\beta^2} 
	          \tanh \bigg(\frac{z_{\rm I}}{l_{\rm FM}}\bigg)    
	          \Big\{\beta-j_{s,{\rm FM}}(z_{\rm I})\Big\}    \bigg] \label{eq:jsia} \\
j_{s,{\rm NM}}(z_{\rm I})&=\gamma-\frac{(1-\gamma^2)\delta}{AR_{\rm I} \sinh \delta}
	\bigg[\rho_{\rm NM}l_{\rm NM}j_{s,{\rm NM}}(z_{\rm I})  \cosh \delta -
	\frac{\rho_{\rm FM} l_{\rm FM}}{1-\beta^2} \tanh\bigg(\frac{z_{\rm I}}{l_{\rm FM}}
	\bigg)    \Big\{\beta-j_{s,{\rm FM}}(z_{\rm I})\Big\}    \bigg]. \label{eq:jsib}
	\end{align}
\end{subequations}	
\end{widetext}
Assuming that we know $\rho_{\rm NM}$ and $l_{\rm NM}$ for NM, $\beta$, $\rho_{\rm FM}$, and $l_{\rm FM}$ for FM and $AR_{\rm I}$ for the interface, then using \eqref{eq:jsi} we can determine $\gamma$ and $\delta$ if we know $j_{s,{\rm NM}}(z_{\rm I})$ and $j_{s,{\rm FM}}(z_{\rm I})$. 

Because of their implicit nature, \eqref{eq:jsi} can only be solved numerically. We will need to determine the sensitivity of the solutions  to uncertainities in all of the parameters as well as the extrapolated values $j_{s,{\rm FM}}(z_{\rm I})$ and $j_{s,{\rm NM}}(z_{\rm I})$. To identify the factors limiting the accuracy with which the parameters and spin-currents can be calculated, we need to recall some aspects of the scattering formalism used to calculate these quantities. This we do in the next section.

\section{Calculations}
\label{sec:Calculations}

Within the framework of density functional theory \cite{Hohenberg:pr64, Kohn:pr65}, we solve the quantum mechanical scattering problem \cite{Datta:95} for a general two terminal $\mathcal{L}|\mathcal{S}|\mathcal{R}$ configuration of a Pt$|$FM$|$Pt scattering region ($\mathcal{S}$) embedded between ballistic left ($\mathcal{L}$) and right ($\mathcal{R}$) Cu leads using a wave-function matching (WFM) method \cite{Ando:prb91} implemented \cite{Xia:prb06, Zwierzycki:pssb08} with a tight-binding (TB) muffin-tin orbital (MTO) basis \cite{Andersen:prl84, *Andersen:85, *Andersen:prb86} and generalized to include spin-orbit coupling and noncollinearity \cite{Starikov:prl10, Starikov:prb18} as well as temperature induced lattice and spin disorder \cite{LiuY:prb11, LiuY:prb15}. The solution yields the scattering matrix $S$, from which we can directly calculate the conductance, as well as the full quantum mechanical wave function throughout the scattering region from which we can calculate position dependent charge and spin currents \cite{WangL:prl16, Wesselink:prb19}. The relevant spin current in this study is $j_{sz}^z(z)$ where the superscript indicates the direction of (charge or spin) transport and the subscript indicates the orientation of the spins which is here the magnetization direction $\widehat{\bf m}$ of Py, chosen to be parallel to the transport direction $z$. In this section, we discuss the considerations we make specifically for the Pt$|$Py$|$Pt and Pt$|$Co$|$Pt scattering region to extract reliable interface parameters. 

\subsubsection{Supercells: Lattice mismatch}

To model various types of disorder, we assume periodicity in the directions transverse to the transport direction and construct periodic ``lateral supercells'' with which to model interfaces between fcc materials like Pt and Py that have different lattice parameters, $a_{\rm Pt}=3.923\,$\AA\ and $a_{\rm Py}=3.541\,$\AA\,, respectively. Bulk Co is typically hcp below roughly 700 K \cite{Betteridge:pms80}. However, when interfaced with a material like Pt, the Co thin films are predominantly fcc \cite{Lamelas:prb89, deGronckel:prb91}. By preserving the volume of hcp Co with lattice parameters $a=2.507\,$\AA\, and $c=4.069\,$\AA, we obtain an effective fcc lattice constant $a_{\rm Co}=3.539\,$\AA. The similarity of this value to the lattice constant of Py means that Co and Py can be treated interchangeably when modelling the interface with Pt.

\begin{figure}[t]
	\centering
	\includegraphics[width=1.0\linewidth]{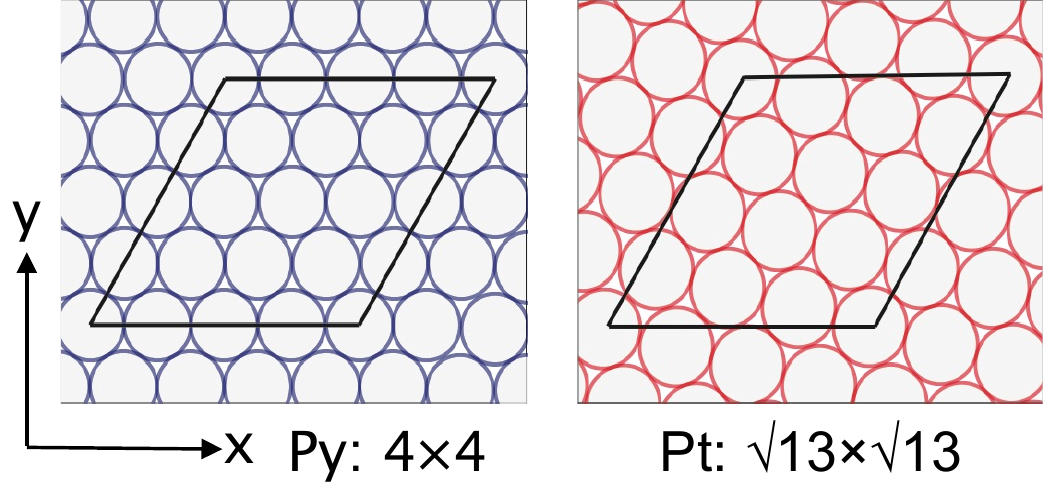}
	\caption{Two dimensional (111) atomic planes of Py (left) and Pt (right). The parallelogram outlines the equivalent unit cells.}
	\label{figC}
\end{figure}

For a given orientation of an A$|$B interface between materials A and B, the periodicity of the atoms in the plane of the interface is described by primitive lattice vectors $\lbrace {\bf a}_1, {\bf a}_2 \rbrace $ and $\lbrace {\bf b}_1, {\bf b}_2 \rbrace$. We construct an ordered list of all A and B in-plane lattice vectors $|n_1 {\bf a}_1 + n_2 {\bf a}_2|$, respectively $|m_1 {\bf b}_1 + m_2 {\bf b}_2|$ and select a pair acceptably close in magnitude. In general, making these coincide will require rotating the two lattices with respect to one another as sketched in \cref{figC} where a 4$\times$4 unit cell of (111) oriented Py (or Co) is matched to a $\sqrt{13}\times\sqrt{13}$ unit cell of similarly oriented Pt. These ``superlattices'' match to better than 0.1\% and the residual mismatch is accommodated by uniformly expanding Pt. To ensure that the spin currents do not depend on the artificial lateral periodicity \cite{Wesselink:prb19}, the unit cells are doubled to 8$\times$8 for FM and $2\sqrt{13}\times2\sqrt{13}$ for Pt. These considerations are subject to the constraint that the computational expense of solving the scattering problem scales as the third power of the number of atoms in a lateral supercell.

The assumption of periodicity transverse to the transport direction allows us to make use of Bloch's theorem to label the wave functions that are solutions of the Schr{\"{o}}dinger equation with a two dimensional (2D) wavevector. The corresponding 2D Brillouin zone (BZ) is sampled with 32 $\times$ 32 $k$-points for these supercells leading to equivalent samplings for 1$\times$1 unit cells of Py and Pt of 256$\times$256 and $\sim$230$\times$230 respectively. 

\subsubsection{Thermal disorder}
\label{sec:thdis}

To carry out finite temperature calculations, we use a frozen thermal disorder scheme \cite{LiuY:prb11, LiuY:prb15, Starikov:prb18} to displace atoms from their equilibrium positions and rotate magnetic moments from their equilibrium orientations. The distribution of the (uncorrelated) atomic displacements is assumed to be Gaussian and is characterized by a root-mean square displacement $\Delta$. For Pt, we choose a value of $\Delta$ to reproduce the experimental resistivity \cite{HCP90, Ho:jpcrd83} at any given temperature. For a given value of $\Delta$, multiple ($\sim$10-20) random configurations of disorder are generated and all calculations are averaged over these configurations. For Py, $\Delta$ is derived from the Debye model \cite{Tanji:jpsj71} and spin disorder is modelled with a Gaussian distribution of rotations to reproduce the experimental magnetization \cite{Wakelin:ppsb53} for a given temperature; this prescription has been shown to reproduce the experimentally observed resistivity very well \cite{LiuY:prb15, Starikov:prb18}. 

For Co, the spin disorder is modelled with a Gaussian distribution of polar rotation angles to reproduce the experimental magnetization \cite{Kuzmin:prl05} for a given temperature. Most experiments report the polycrystalline resisitivity profile for Co as a function of temperature \cite{Betteridge:pms80}. Masumoto et al. \cite{Masumoto:jjimm66} measured the experimental resistivity for single crystal hcp Co at 300 K. We choose a value of $\Delta$ so that together with the spin disorder, the experimental resistivity at room temperature along [0001], $\rho_{\rm Co}=10.28~\mu\Omega{\rm cm}$ \cite{Masumoto:jjimm66} is reproduced. Since there is no data available for monocrystalline Co at other temperatures and we use fcc instead of hcp Co, we use the Debye model and determine a Debye temperature (450 K) that yields the chosen $\Delta$ at 300 K. $\Delta$ and the corresponding resistivities at all temperatures for fcc Co are then determined using a combination of Debye model with Debye temperature 450 K to describe the lattice disorder and spin disorder that reproduces the experimental magnetization.

Though the resistivity of bulk Pt can be calculated entirely from first principles within the adiabatic approximation by performing first-principles phonon calculations, populating the resulting phonon modes at a fixed temperature $T$, taking snapshots of the superimposed phonons, determining the resistance $R$ of various lengths $L$ of thermally disordered material and finally extracting $\rho$ from  $R(L)$, the agreement with experiment, though good, is not perfect \cite{LiuY:prb15}. Spin disorder in magnetic materials can be treated analogously but additional approximations are necessary because spin-wave theory underestimates the magnetization decrease induced by temperature \cite{LiuY:prb15}. The tediousness of calculating phonon and magnon dispersion relations for magnetic alloys motivated us to adopt the simpler Gaussian disorder approach sketched above not only for Py but also for Pt and Co. Thus in the results we will present below, experimentally observed bulk resistivities are reproduced by construction.

\subsubsection{Potentials}

Bulk potentials for all atomic species (Cu, Pt, Ni, Fe, Co) are calculated in the atomic spheres (AS) approximation (ASA) using the TB-LMTO method  \cite{Andersen:prl84, *Andersen:85, *Andersen:prb86}. AS potentials for Ni and Fe are evaluated self-consistently for the fcc substitutional random alloy Py using the coherent potential approximation (CPA) \cite{Soven:pr67} implemented with TB-MTOs \cite{Turek:97}. 

In many experiments involving (Pd and) Pt, interface effects are expected to depend strongly on proximity-induced Pt magnetization \cite{Huang:prl12}. Since we are focussing on the evaluation of interface parameters in this paper, we will test this hypothesis by constructing interfaces with and without proximity-induced magnetism. In the simplest, default scenario, no magnetism is induced in Pt by proximity to FM. Because of the complexity of the FM$|$Pt interface and the inability of current CPA implementations to treat large unit cells, we study the magnetic moments induced in Pt by a putative FM grown pseudomorphically on ``bulk'' Pt; the in-plane lattice constant of a (111) oriented fcc FM is expanded to match that of Pt while the out-of-plane lattice constant is reduced to keep the volume of the FM unchanged. 

\begin{figure}[t]
\includegraphics[width=1\linewidth]{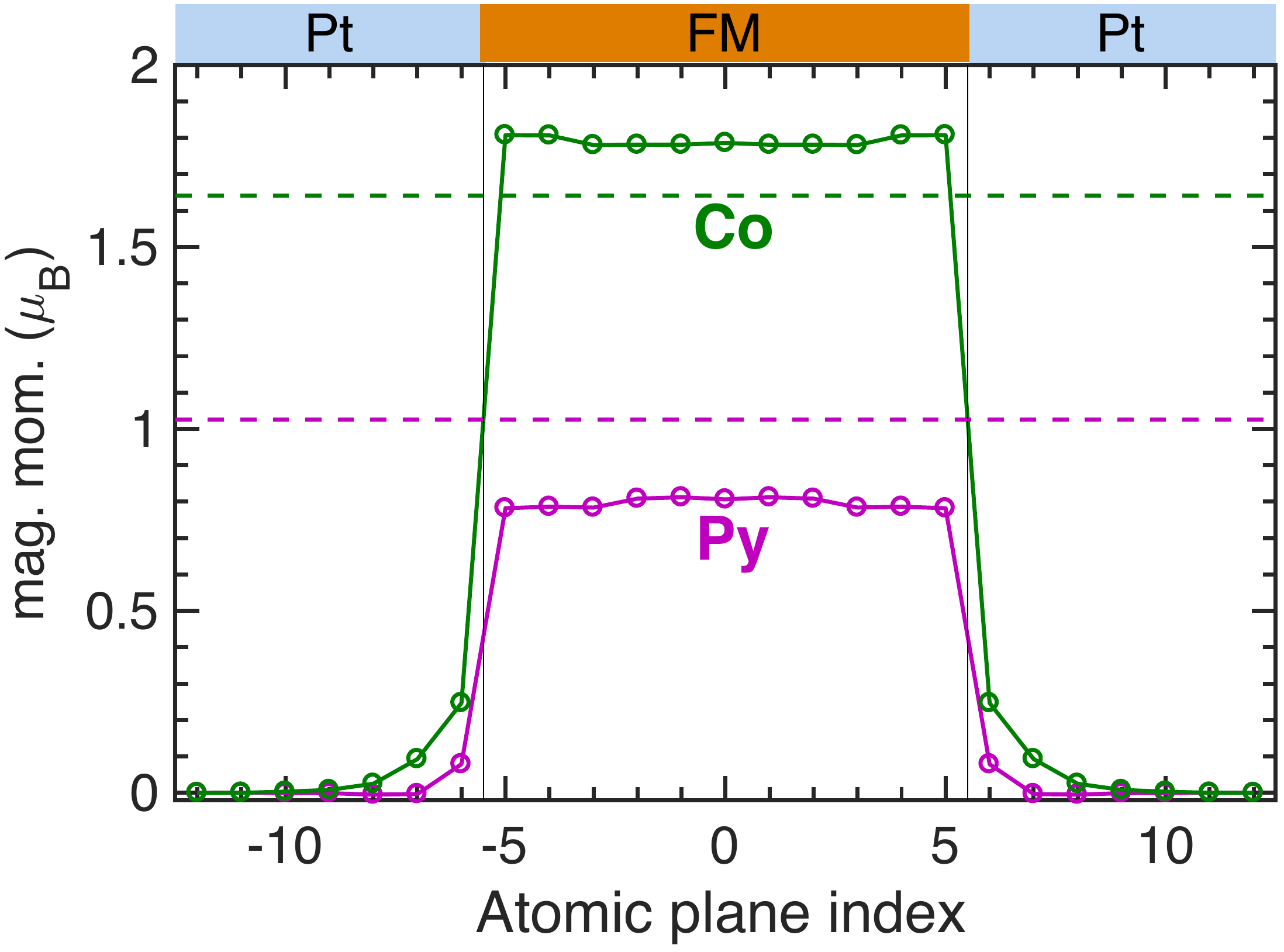}
\caption{Magnetic moment profile of an fcc (111) oriented Pt$|$FM$|$Pt (FM=Py, Co) geometry calculated self-consistently using the coherent potential approximation. 11 layers of trigonally distorted pseudomorphic FM are sandwiched between semiinfinite Pt. The atomic moment is shown as a function of the plane number for Py (pink) and Co (green). For comparison, the horizontal dashed lines indicate the atomic moments calculated for bulk unstrained fcc Co (green) and Py (pink).}
\label{figD}
\end{figure}

The magnetization profiles of 11 atomic layers of pseudomorphic Py and Co sandwiched between cubic fcc Pt obtained with self-consistent CPA calculations are shown in \cref{figD}. The magnetic moment induced in Pt by proximity to Py decreases rapidly from 0.08~$\mu_B$ per Pt atom in the interface layer to 0.005~$\mu_B$ (by 95\%) over 5 layers. The magnetic moment profile of Py is essentially constant at 0.8~$\mu_B$ per atom with little change in the interface layer. For comparison, the average magnetic moment of bulk unstrained Py is 1~$\mu_B$ so that the effect of the 10.7\% strain is to reduce the Py moment. 

Bulk fcc Co has a large magnetic moment of 1.64~$\mu_B$. Stretching its in-plane lattice constant $a_{\rm Co}=3.55 \,$\AA\ to make it match $a_{\rm Pt}$ while conserving its volume increases the moment to 1.78~$\mu_B$. This stretched Co induces a moment of 0.25~$\mu_B$ in the adjacent Pt layer. The induced moment decreases to 0.003~$\mu_B$ (by 99\%) over 5 layers. Why strain affects the moments in Py and Co differently goes beyond the scope of this publication.

In the scattering calculations to be discussed in the next section, we will replace the bulk non-magnetic Pt potentials for 7 layers of Pt next to the interface with magnetized Pt potentials and compare the resulting spin currents in the two cases. 
Since the Py moment is reduced by strain in the pseudomorphic structure and this may result in an underestimation of the moments induced in Pt, we will use the spin polarized Pt potentials determined for Pt$|$Co$|$Pt as input to the scattering calculations for both Py$|$Pt and Co$|$Pt interfaces. 

To determine the spin-flip diffusion length, we will in the next section inject a fully spin-polarized current $(|j_s|=1)$ into the scattering region from ``half-metallic ferromagnetic'' (HMF) Cu leads denoted Cu$\uparrow$ or Cu$\downarrow$   \cite{Starikov:prb18, Wesselink:prb19}. These artificial leads are constructed by adding a constant repulsive term to the AS potential of the minority (majority) spin states so that these states are lifted above the Fermi level and the spin-current density consequently consists of only majority (minority) spin states.

\subsubsection{Slab length}

Eqs.~\eqref{eq:jsi} include an apparent dependence on the slab length $z_{\rm I}=L_{\rm FM}/2$. While it is expected that the spin current must saturate to $\beta$ at the center of a sufficiently long FM slab, it is not a priori clear how the spin current close to the interface depends on the slab length. It turns out that for $L_{\rm FM}\ge 6 l_{\rm FM}$, the spin current close to the FM$|$NM interface is independent of $L_{\rm FM}$. Both left and right slabs of Pt have $L_{\rm Pt}> 4 l_{\rm Pt}$) to ensure any spin current in Pt has decayed to a negligible value close to the leads.

\section{Results and Discussion}
\label{sec:Results}

The bulk transport properties of Pt, Py and Co are calculated for the temperature range 100-500 K in \cref{sec:BM} allowing $j_{s,{\rm FM}}(z_{\rm I})$ and $j_{s,{\rm Pt}}(z_{\rm I})$ to be determined. In \cref{sec:IR} we calculate the interface resistance $AR_{\rm I}$ for an FM$|$Pt interface using the Landauer-B{\"{u}}ttiker formalism. The remaining two parameters $\gamma$ and $\delta$ are determined by solving equations \eqref{eq:jsi}. No calculations were carried out at 100 K for bulk Co or Pt$|$Co$|$Pt because  $l_{\rm sf}^{\rm Co}$ was estimated to be $\ge 25 \,$nm which when combined with the calculated value of $l_{\rm sf}^{\rm Pt}=22 \,$nm would require constructing a trilayer with more than 100,000 atoms and require excessive computational resources.

\subsection{Bulk materials}
\label{sec:BM}

The bulk resistivities $\rho_{\rm Pt}$, $\rho_{\rm Py}$ and $\rho_{\rm Co}$ can be determined directly from the scattering matrix using the Landauer-B{\"{u}}ttiker formalism. The transport polarization $\beta$ for Py and Co as well as the spin-flip diffusion lengths $l_{\rm Pt}$, $l_{\rm Py}$ and $l_{\rm Co}$ are extracted from calculations of the spin current $j_s(z)$. Using these bulk parameters we fit the $j_s(z)$ calculated for Pt$|$Py$|$Pt and Pt$|$Co$|$Pt trilayers shown in \cref{figB} (top) and (bottom), respectively. Fitting $j_s(z)$ in the FM material yields the orange curve in \cref{figB} and the value of $A_{\rm FM}$ using \eqref{eq:jFM}. 
Fitting $j_s(z)$ in Pt yields the blue curve in \cref{figB} and the value of  $B_{\rm Pt}$ using \eqref{eq:jNM}. This allows us to calculate $j_{s,{\rm FM}}(z_{\rm I})$ and $j_{s,{\rm Py}}(z_{\rm I})$ by extrapolation to the interface.

\subsubsection{Resistivity}
\label{sssec:res}

\begin{figure}[t]
\includegraphics[width=8.6 cm]{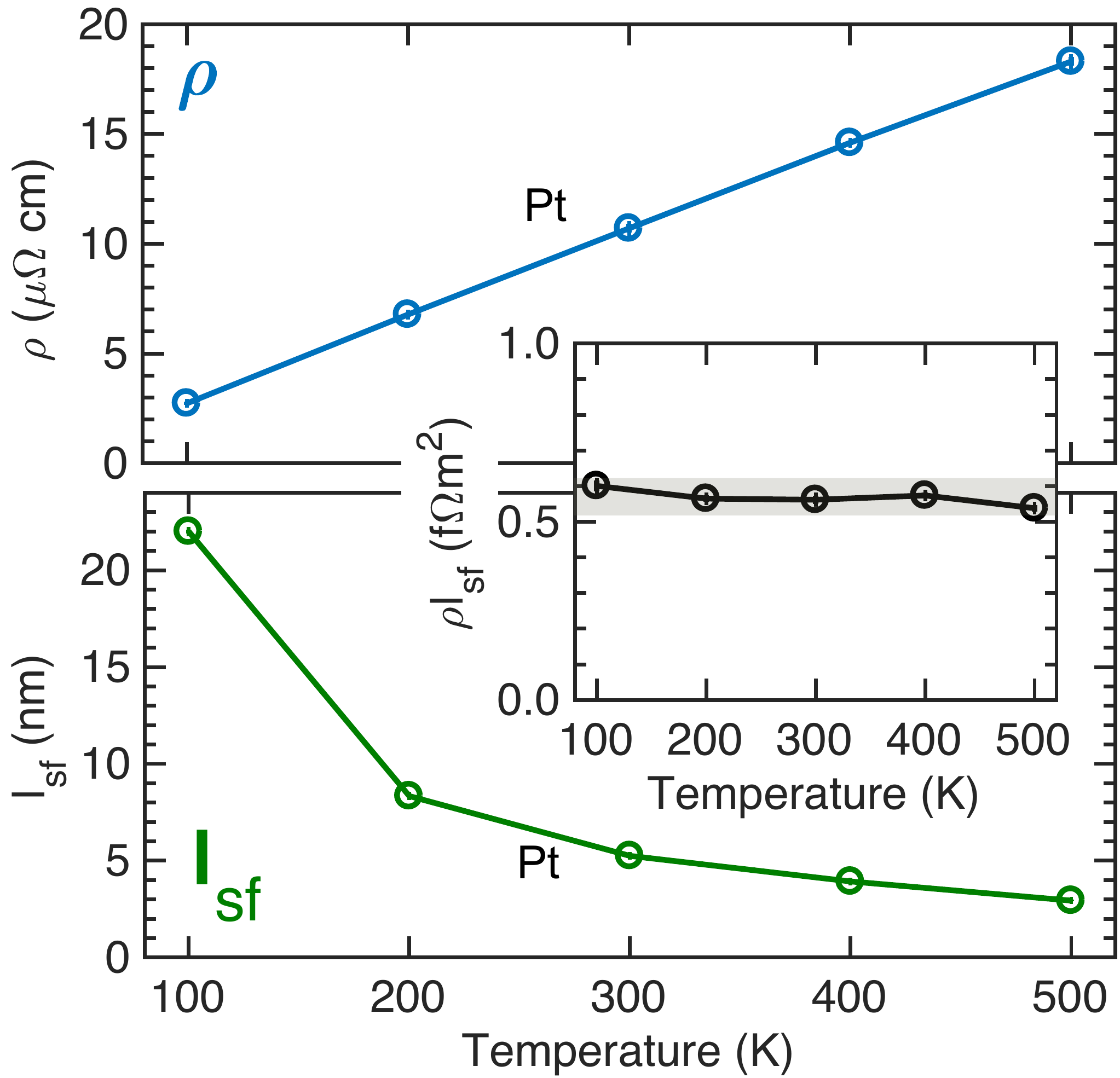}
\caption{Resistivity $\rho_{\rm Pt}$ and spin-flip diffusion length $l_{\rm Pt}$ as a function of temperature for bulk Pt. The product $\rho l_{\rm sf}$ is shown in the inset. }
\label{figE}
\end{figure}

\begin{figure}[t]
\includegraphics[width=8.6 cm]{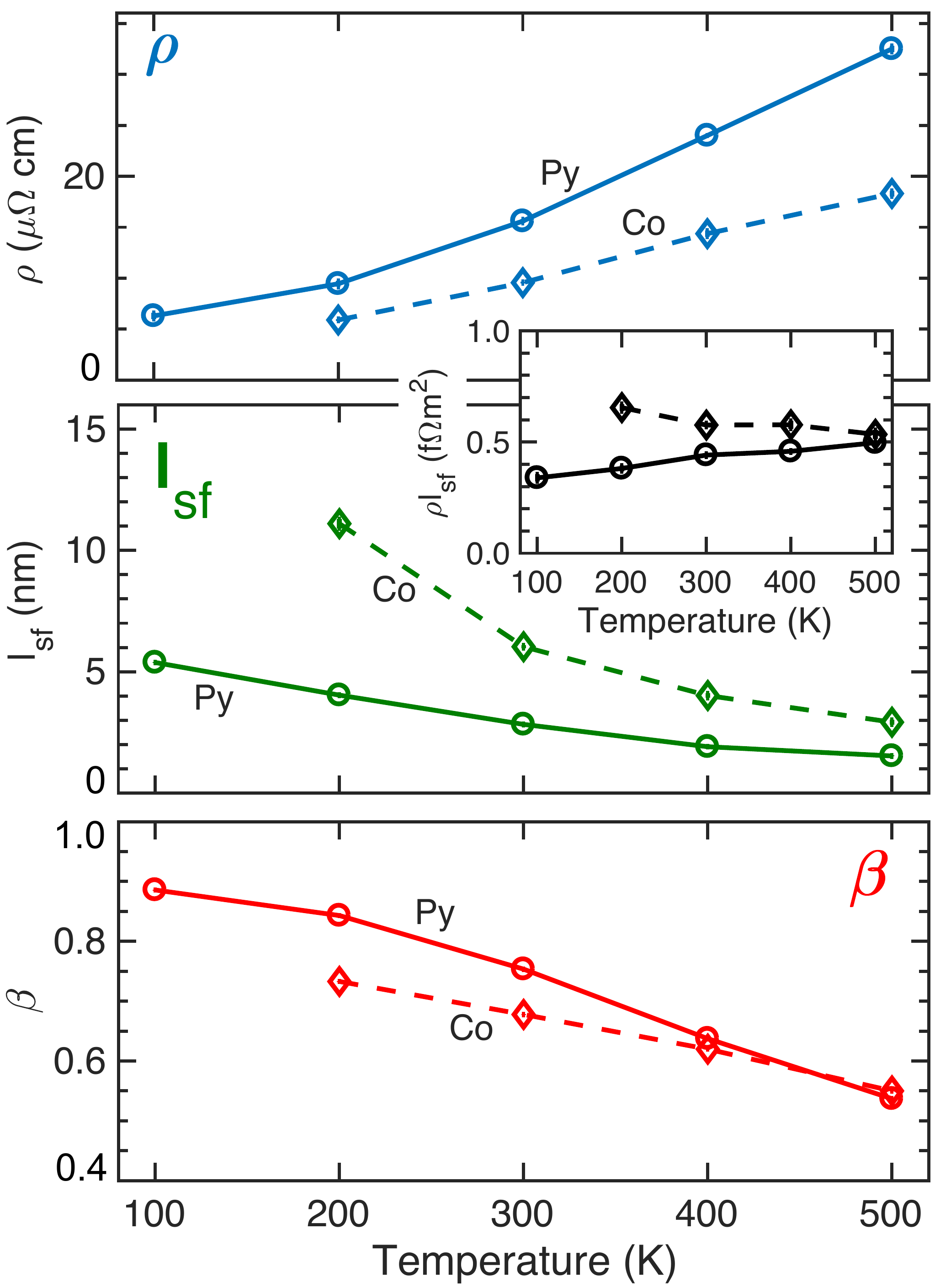}
\caption{Resistivity $\rho$, spin-flip diffusion length $l_{\rm sf}$ and spin-asymmetry parameter $\beta$ as a function of temperature for bulk Py (circles) and bulk Co (diamonds). The product $\rho l_{\rm sf}$ is shown in the inset.}
\label{figG}
\end{figure}

By construction, the Gaussian lattice and spin disorder we use reproduces the experimentally observed \cite{HCP90, Ho:jpcrd83} resistivities of bulk Pt (\cref{figE}) and Py (\cref{figG}) at all temperatures. For Co, the lattice and spin disorder chosen at 300 K as described in \cref{sec:thdis} reproduce the experimental resistivity value of $10.28~\mu\Omega{\rm cm}$ observed for crystalline hcp Co \cite{Masumoto:jjimm66}. Using the same lattice and spin disorder for an fcc structure yields a resistivity of $9.56\pm0.08~\mu\Omega{\rm cm}$ at 300 K. Co fcc resistivities at other temperatures are calculated as described in \cref{sec:Calculations} and are plotted in \cref{figG}. Here, it is important to note how the resistivity behaves for Py and Co to better anticipate the behaviour of the remaining spin transport parameters. The chemical, alloy disorder in Py leads to a finite value of $\rho \sim4\mu\Omega~{\rm cm}$ at 0 K whereas the resistivity of ordered Co approaches zero. A Curie temperature of 872~K for Py versus 1385~K for Co implies that at any given temperature the magnetic ordering in Co is stronger. The combination of chemical and magnetic disorder in Py makes its resistivity change more rapidly as a function of temperature.  

All other parameters that we report have been calculated using the same thermal disorder employed for the resistivity calculations. We did not attempt to reproduce the resistivities reported for thin films that differ from the known bulk values because so little is known about the microscopic disorder (impurities, vacancies, self interstitials, grain boundaries, surfaces etc.) that might give rise to the differences. 

\subsubsection{Spin-flip diffusion length $l_{\rm sf}$: Pt}

A fully polarized spin current $j_s(0)=1$ injected into thermally disordered Pt decays exponentially to zero, $j_s(z)=C \exp(-z/l)$. Using the procedure described by Wesselink {\it et al.} \cite{Wesselink:prb19} for a Au$\uparrow$$|$Pt$|$Au scattering geometry, we obtain $l_{\rm Pt}$ from spin current calculations for temperatures between 100-500 K. The results for $l_{\rm Pt}$ shown in \cref{figE} exhibit a 1/T dependence and, as shown in the inset, satisfy the relationship $\rho_{\rm Pt}l_{\rm Pt}=0.57\pm0.05 \,$f$\Omega$$m^2$ \cite{Nair:prl21} in accordance with the Elliott-Yafet mechanism that is based upon free-electron like energy dispersion \cite{Elliott:pr54, Yafet:63}.

%

\subsubsection{$l_{\rm sf}$ and $\beta$: Py and Co}

\begin{figure}[t]
\includegraphics[width=8.6 cm]{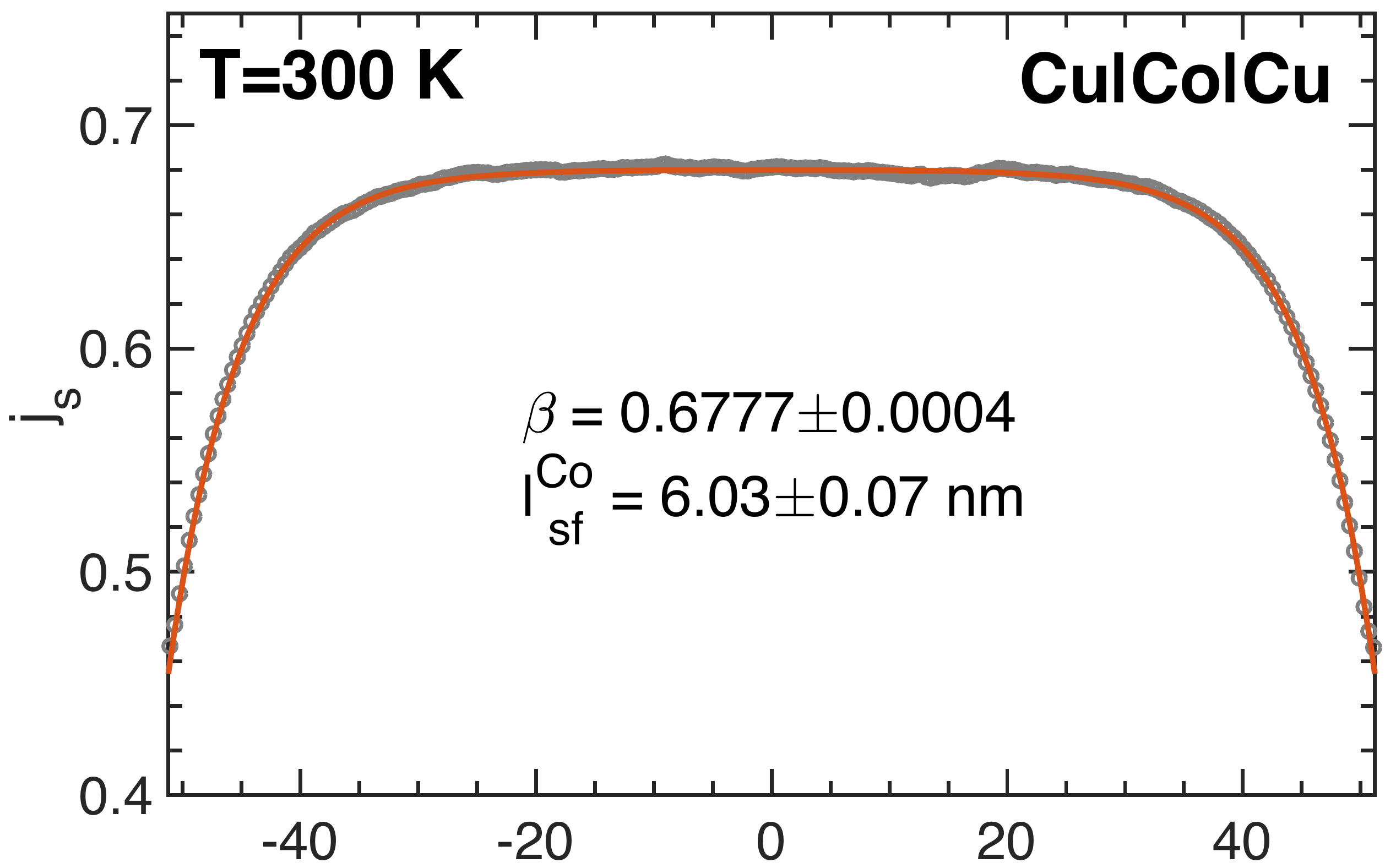}
\caption{An unpolarized charge current enters a sufficiently long slab of an 8$\times8$ lateral supercell of thermally disordered Co at 300 K. The grey circles show the resulting spin current $j_s$ polarized along the magnetization direction $\hat{z}$. The orange curve shows the fit for $j_s$ that yields $\beta=0.6777\pm0.0004$ and $l_{\rm sf}=6.03\pm0.07$ nm.}
\label{figF}
\end{figure}

A charge current passed through a magnetic material is naturally polarized along the magnetization direction. The way in which it approaches its equilibrium polarization value $\beta$ is described in equation \eqref{eq:jFM} by the spin-flip diffusion length $l_{\rm FM}$. Wesselink {\it et al.} studied the computational aspects of extracting $\beta$ and $l_{\rm Py}$ from $j_s(z)$ calculated for Py at 300 K in Ref.~\onlinecite{Wesselink:prb19}. Here we show the $z$ dependence of the spin current calculated for a symmetric Cu$|$Co$|$Cu scattering geometry with room temperature thermal disorder in \cref{figF}. By fitting $j_s(z)$ to \eqref{eq:jFM}, we extract values of $\beta=0.68$ and $l_{\rm Co}=6.03 \,$nm for room temperature Co. 

The temperature dependence of $l_{\rm FM}$ and $\beta$ is plotted in \cref{figG} for both Py and Co.  $l_{\rm Py}$ decreases from 5.4 nm at 100 K to 1.5 nm at 500 K. Co has a much larger SDL, $l_{\rm Co}=11.1$ nm at 200 K that decreases to 2.9 nm at 500 K. The smaller values of $l_{\rm FM}$ for Py can be attributed to the chemical disorder that is present at all temperatures in addition to the thermal spin and lattice disorder. 
Unlike Pt that conforms to the behaviour predicted by the Elliott-Yafet model in spite of not having free-electron like energy bands, the product $\rho l_{\rm sf}$ is not a constant at all temperatures for either Py or Co, as shown in \cref{figG} (inset).
With thermal spin disorder in addition to the lattice disorder of Pt, $\rho_{\rm Co} l_{\rm Co}$ decreases with increasing temperature. Py has in addition alloy disorder and $\rho_{\rm Py} l_{\rm Py}$ exhibits the opposite behaviour. 
The interplay of thermal (lattice and spin) and chemical disorder combined with the complex $d$ electron band structure does not allow us to provide a simple picture with which to explain these findings.

The polarization $\beta$ is proportional to the difference between the minority and majority resistivities. For Py at 0 K, the minority resistivity $\rho_{\rm min}$ is $\sim$200 times larger than the majority resistivity $\rho_{\rm maj}$ \cite{Starikov:prb18} giving rise to a value of $\beta \sim 0.99$. This large value decreases to still large values of 0.88 at 100 K, 0.75 at RT and 0.54 at 500 K \cite{LiuY:prb15}. Below 400~K, $\beta_{\rm Co}$ is smaller than $\beta_{\rm Py}$ and the difference can be understood as follows. The majority-spin potential of Co is very similar to the majority-spin potentials of Fe and Ni in Py, all of which have fully occupied majority-spin 3$d$ bands. The very small, majority-spin resistivity of Co is therefore comparable to that of Py, $\rho_{\rm maj}^{\rm Co} \sim \rho_{\rm maj}^{\rm Py}$ in which case $\beta_{\rm Co} < \beta_{\rm Py}$ implies that $\rho_{\rm min}^{\rm Co} < \rho_{\rm min}^{\rm Py}$ because the minority-spin electrons in Co are not scattered by alloy disorder as in Py; the RT bulk fcc resistivities are $\rho_{\rm Co}=9.6~\mu\Omega{\rm cm}$ versus $\rho_{\rm Py}=15.6~\mu\Omega{\rm cm}$. At very low temperatures, in the ballistic regime, resistivities must be replaced by resistances and $\beta_{\rm Co} \sim -0.45$ because the minority-spin Sharvin conductance of fcc Co is larger than its majority-spin conductance \cite{Schep:prb98}. Thus, we predict a change in the sign of $\beta_{\rm Co}$ as a function of temperature below 200~K. Above 400~K, temperature-induced spin disorder in Py significantly lowers the polarization, which is nearly the same as that of Co.

\subsection{Interfaces}
\label{sec:Ints}

\begin{figure}[t]
\includegraphics[width=8.6 cm]{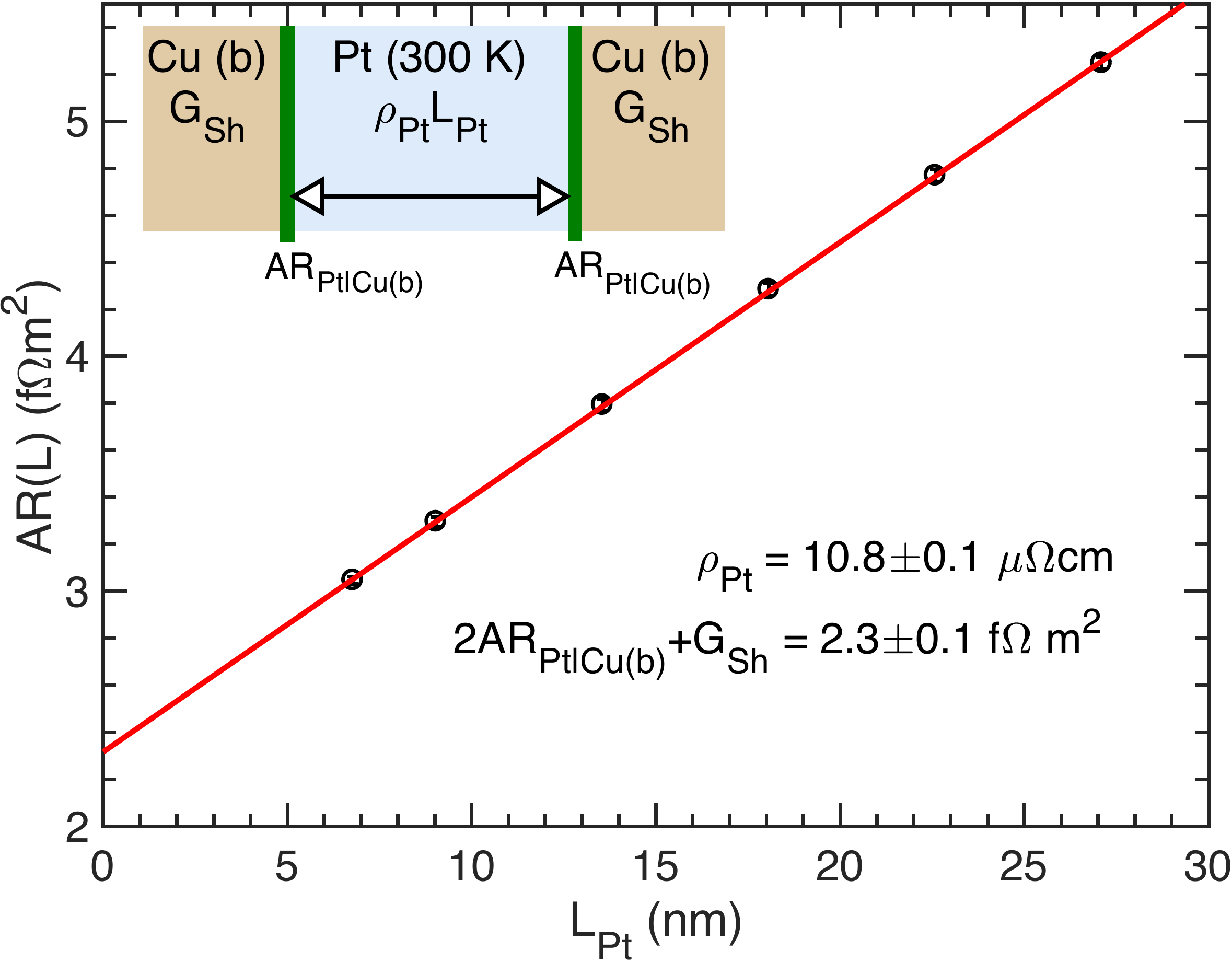}
\caption{Total resistance of a length $L_{\rm Pt}$ of diffusive Pt sandwiched between ballistic Cu leads as a function of $L_{\rm Pt}$. A linear fit $R(L)$ yields $\rho_{\rm Pt}$ as the slope; the intercept is a sum of interface and Sharvin contributions. }
\label{figH1}
\end{figure}

Before calculating the interface resistance $AR_{\rm I}$ for e.g., an FM$|$Pt interface using the scattering formalism, we need to reexamine how the resistivity of Pt was calculated in \cref{sssec:res}. Using the Landauer-B\"{u}ttiker formalism \cite{Datta:95}, we calculated the conductance $G$ of a scattering region containing thermally disordered Pt, expressing $G$ in terms of the probability that Bloch states in the left lead $\mathcal{L}$ attached to the scattering region $\mathcal{S}$ are transmitted through the scattering region into the right lead $\mathcal{R}$. The result of doing this for a length $L$ of thermally disordered Pt sandwiched between ballistic Cu leads is shown in \cref{figH1} for $T=300\,$K. The resistivity was extracted as a slope of the linear fit of $R(L)=1/G(L)$ plotted as a function of $L$ \cite{Starikov:prl10, LiuY:prb11, Starikov:prb18}. 

However, for $L=0$, the resistance does not vanish; there is a finite intercept because a finite cross section $A$ of a ballistic material has a finite conductance, the Sharvin conductance $G_{\rm Sh}$ \cite{Sharvin:zetf65}. In addition, there is a resistance $R_{\rm Cu|Pt}$ associated with each interface so the total resistance consists of  
\begin{equation}
AR(L_{\rm Pt})=\rho_{\rm Pt}L_{\rm Pt} +2AR_{\rm Cu|Pt} + 1/G_{\rm Sh}.
\label{eq:13} 
\end{equation}
This interface resistance, however, is for an interface between a ballistic ($T=0$) Cu lead and diffusive Pt. To determine the interface resistance $AR_{\rm I}$ between two diffusive interfaces requires more work. 

\subsubsection{Interface resistance}
\label{sec:IR}

\begin{figure}[t]
\includegraphics[width=8.6 cm]{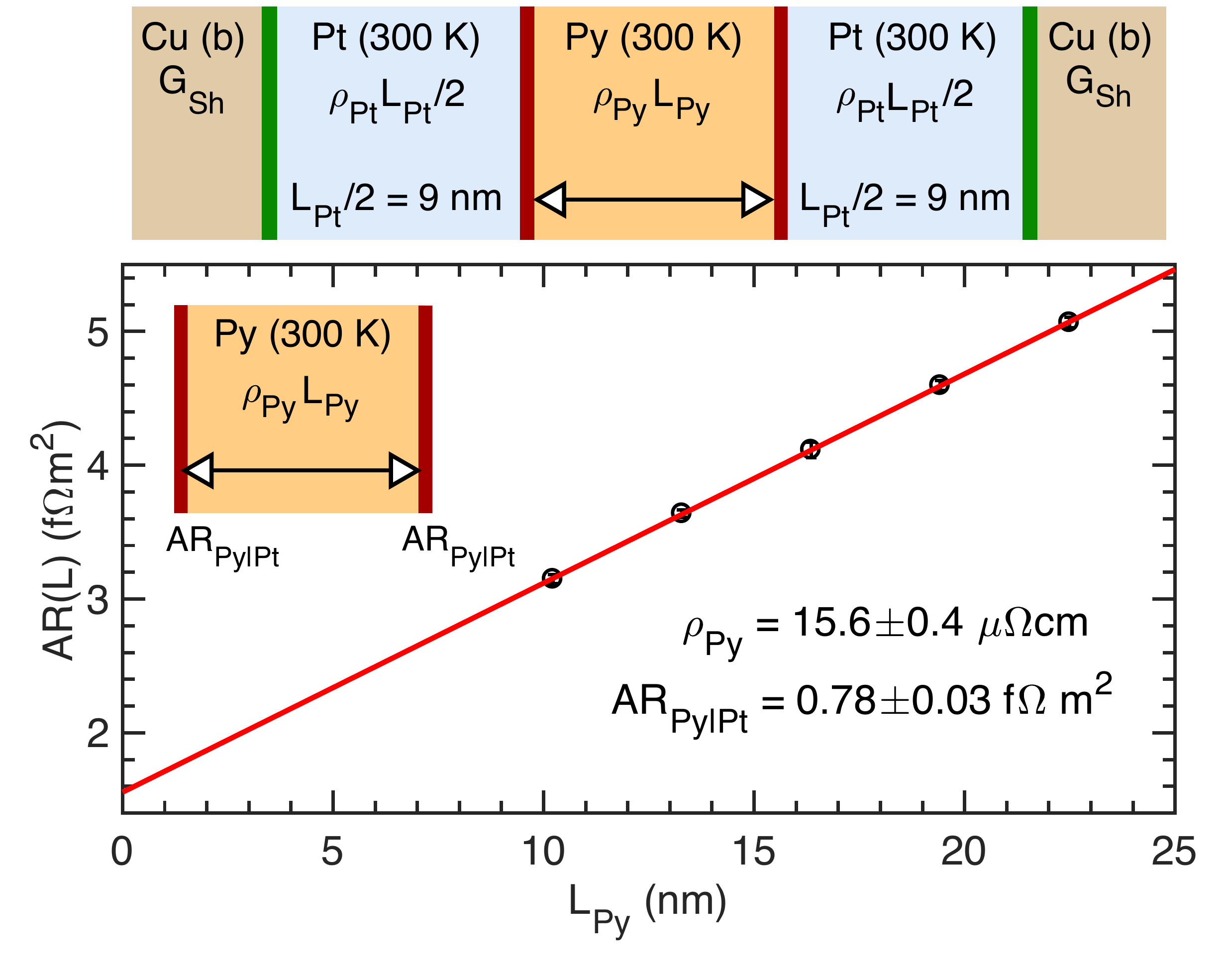}
\caption{Total resistance of a diffusive Pt$|$Py$|$Pt trilayer sandwiched between Cu leads as a function of the Py thickness $L_{\rm Py}$ for a fixed Pt thickness $L_{\rm Pt}= 18\,$nm. To extract $\rho_{\rm Py}$ and $AR_{\rm I}$, resistances for Pt$|$Py$(L_{\rm Py})|$Pt are calculated and all contributions from Cu and Pt are subtracted leaving $AR_{\rm I}$ as the intercept and $\rho_{\rm Py}$ as the slope for the linear fit.}
\label{figI}
\end{figure}

We extract $AR_{\rm I}$ in a two step procedure. We first calculate the total resistance for a symmetric, diffusive Pt$|$FM$|$Pt trilayer embedded between ballistic leads for a variable length $L_{\rm FM}$ of FM and fixed length $L_{\rm Pt}$ of Pt. Both $L_{\rm FM}$ and $L_{\rm Pt}$ should be much longer than the respective mean free paths so that the total areal resistance for the scattering region can be expressed in terms of a series resistor model as
\begin{eqnarray}
AR(L_{\rm FM})&&=\rho_{\rm FM}L_{\rm FM}+\rho_{\rm Pt}L_{\rm Pt} \nonumber \\
&&+2AR_{\rm FM|Pt}+2AR_{\rm Pt|{lead}}+1/G_{\rm Sh}.
\label{eq:LNFNL}
\end{eqnarray}
Here, $R_{\rm FM|Pt}$ is the interface resistance $R_{\rm I}$ we are interested in, $R_{\rm Pt|lead}$ is the interface resistance between Pt and the ballistic lead, and $G_{\rm Sh}$ is the Sharvin conductance of the lead. Noting that 
\begin{equation}
AR(L_{\rm FM}=0)= \rho_{\rm Pt}L_{\rm Pt} + 2AR_{\rm Pt|{lead}}+1/G_{\rm Sh}
\label{eq:LNL}
\end{equation}
we can subtract \eqref{eq:LNL} from \eqref{eq:LNFNL} to obtain
\begin{equation}
A\big[R(L_{\rm FM})-R(L_{\rm FM}=0)\big]= \rho_{\rm FM}L_{\rm FM} + 2AR_{\rm FM|Pt}
\end{equation}
and observe that $AR_{\rm FM|Pt}$ can be calculated as the intercept of the above resistance difference determined as a function of $L_{\rm FM}$. The second step \eqref{eq:LNL} requires calculating the total resistance of a length $L_{\rm Pt}$ of diffusive Pt sandwiched between the same leads used in the first step, as in \cref{figH1} and \eqref{eq:13}.

The result of doing this for RT Py$|$Pt is shown in \cref{figI}. It can be seen that an interface resistance can be determined with an acceptably small uncertainty, $AR_{\rm I} = 0.78 \pm 0.03 \,{\rm f}\Omega {\rm m}^2$. For RT Co$|$Pt, $AR_{\rm I}$ is somewhat larger, $0.85 \pm 0.03 \,{\rm f}\Omega {\rm m}^2$. 

\subsubsection{Spin memory loss and interface spin-asymmetry $\gamma$}

\begin{figure}[t]
\includegraphics[width=8.6 cm]{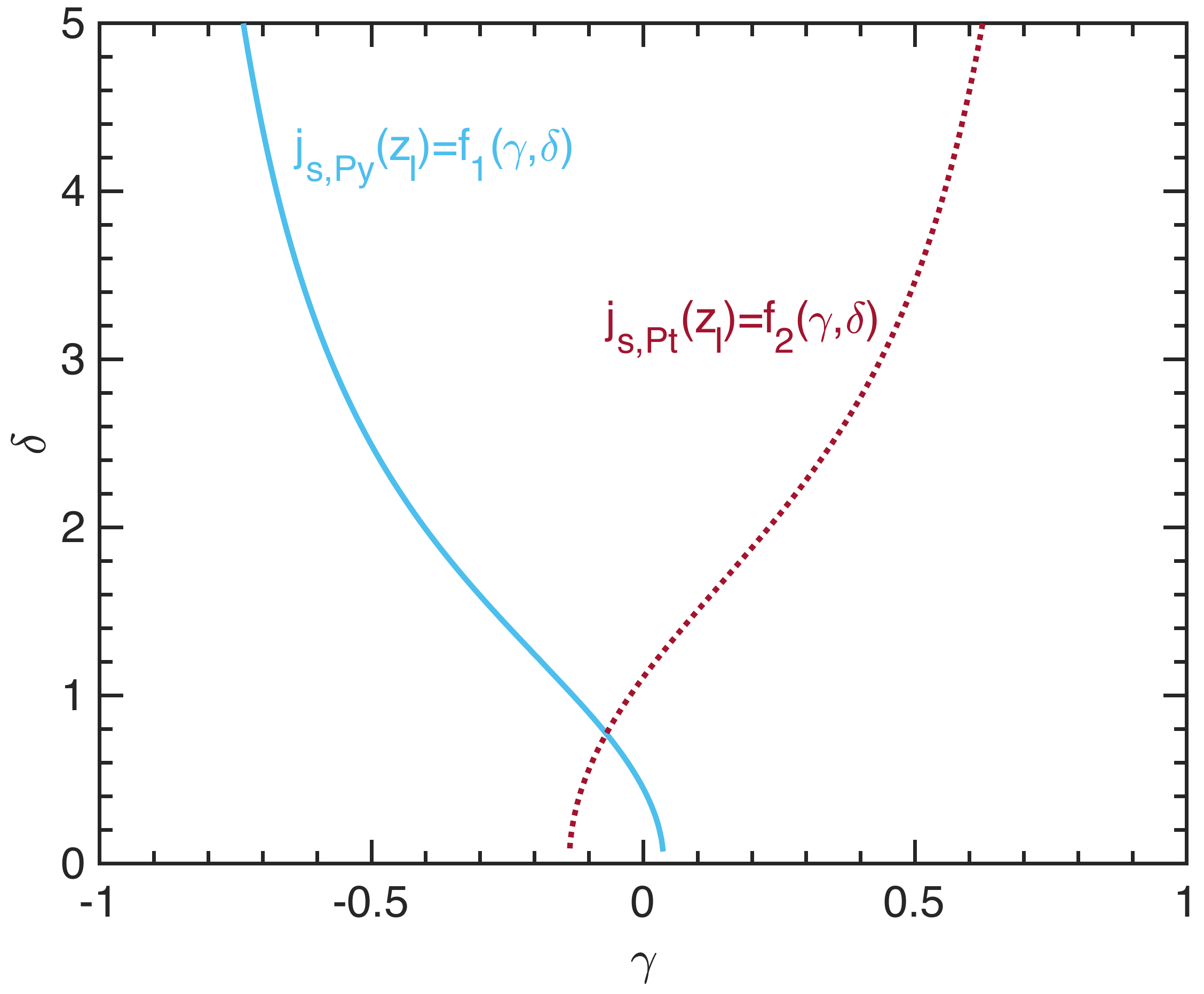}
\caption{Graphical representation of numerical solution for $\gamma$ and $\delta$ obtained by solving \eqref{eq:jsi} for boundary values $j_{s,{\rm Py}}(z_{\rm I})$ and $j_{s,{\rm Pt}}(z_{\rm I})$.}
\label{figJ}
\end{figure}

Now that we know all of the variables besides $\delta$ and $\gamma$ in \eqref{eq:jsi} for both Py$|$Pt and Co$|$Pt, we can solve these two equations simultaneously to yield the remaining two unknown interface parameters. We illustrate this procedure for the RT Py$|$Pt case by substituting the five bulk parameters for Py and Pt as well as the Py$|$Pt interface resistance we have just calculated together with the values of $j_{s,{\rm Pt}}(z_I)$ and $j_{s,{\rm Py}}(z_I)$ into \eqref{eq:jsi}. The equations are graphically represented as contours in $(\gamma,\delta)$ space in \cref{figJ} and the solutions found using standard root-searching algorithms. The single crossing indicates that there exists a unique solution for these parameters. The topology of the crossing indicates the robustness of the solution set. The error bars on these parameters are determined as follows: All input parameters, the five bulk parameters for FM and Pt, $AR_{\rm I}$ and the values of $j_{s,{\rm Pt}}(z_I)$ and $j_{s,{\rm FM}}(z_I)$ span a finite range described by their error bars. Solutions for $\delta$ and $\gamma$ are extracted by substituting all possible combinations of the input parameters into \eqref{eq:jsi}. The range of $\delta$ and $\gamma$ determined from this exercise yields the error bars.

\subsubsection{Temperature dependence of the interface parameters}

\begin{figure}[t]
\includegraphics[width=8.6 cm]{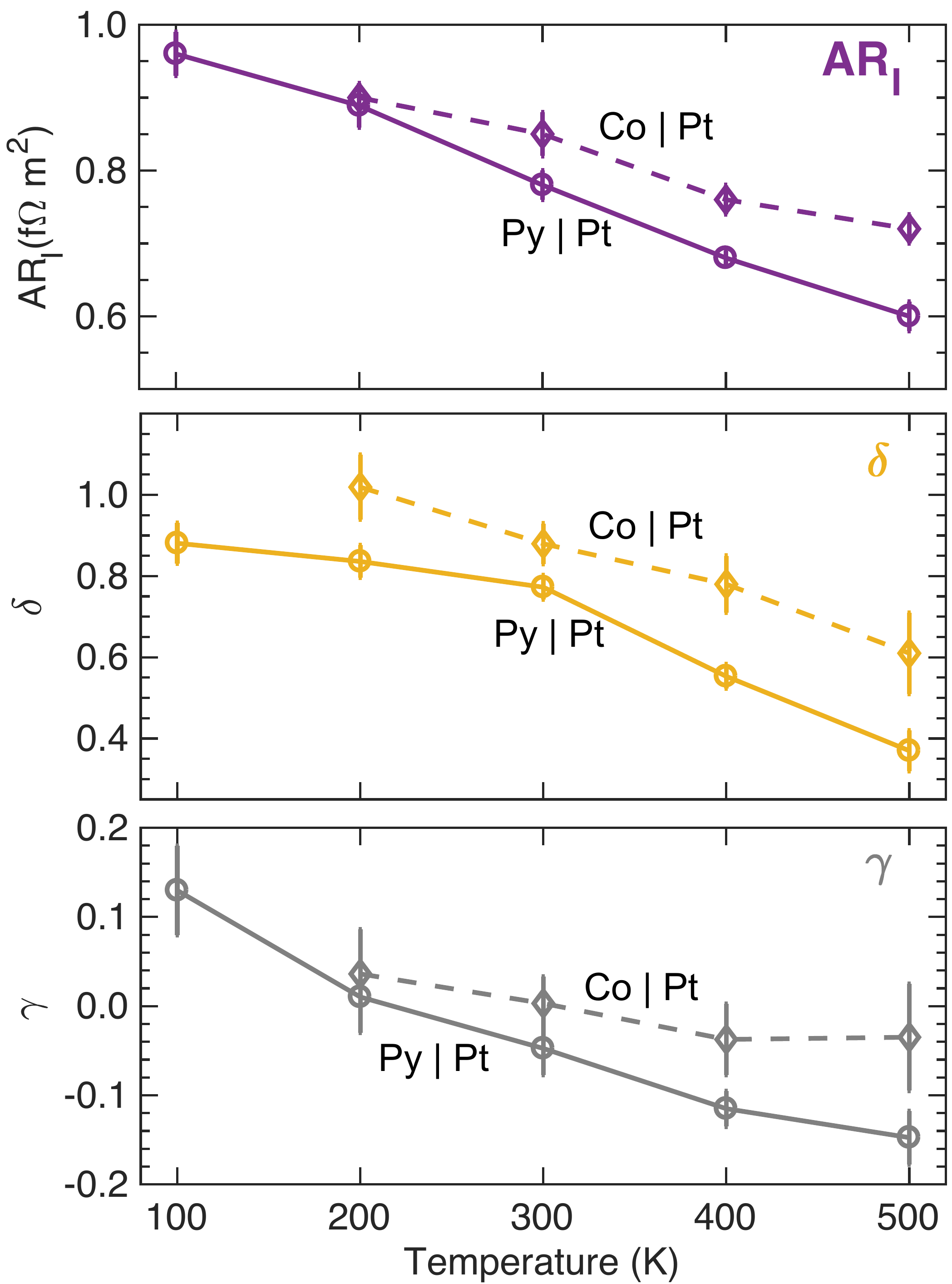}
\caption{Interface parameters $AR_{\rm I},\delta$ and $\gamma$ for Py$|$Pt (circles, solid lines) and Co$|$Pt (diamonds, dashed lines) interfaces plotted as a function of temperature.}
\label{figK}
\end{figure}

The temperature dependence of the interface parameters is shown in \cref{figK}. All parameters are seen to decrease monotonically with temperature for both Py$|$Pt and Co$|$Pt. For an ideal interface between two ballistic solids, the only scattering that occurs is at the interface and the effect of (interface) disorder can be to increase or decrease the transmission through ideal interfaces; this depends strongly on the Fermi surfaces and can be reduced (e.g. Cu$|$Co) or increased (e.g. Fe$|$Cr) by disorder \cite{Xia:prb01}. As the temperature is increased, more scattering occurs in the bulk of the solids so that in the high temperature limit, the relative importance of the interface is reduced. 
In the Landauer-B{\"{u}}ttiker formalism, the transmission probability is proportional to the conductance or inversely proportional to the resistance so the reduction of the interface resistance with increasing temperature seen in \cref{figK} is interpreted as an increased transmission. In  the next section we will see that this increased transmission is dominated by spin disorder.

Since the magnetic ordering in Py is weaker than in Co, the decrease in $AR_{\rm I}$ with temperature is more rapid for Py$|$Pt. $\gamma$ is found to vary in a small range between about -0.15 and 0.15 for Py$|$Pt and between -0.03 and 0.03 for Co$|$Pt, displaying an at best weak correlation with its bulk counterpart $\beta$ in Py and Co. However, $\delta$, the main focus of our interest, shows a significant dependence on temperature and choice of FM. For both interfaces, it decreases monotonically with temperature. Its magnitude is larger for Co$|$Pt compared to Py$|$Pt for all temperatures in the range 200-500 K. At high temperatures, Co disorders more slowly than Py because of its higher Curie temperature, the interface is more abrupt and $\delta$ is higher. We expect the same to hold true at lower temperatures where the interface involving Co  becomes more abrupt than that involving Py as Co orders completely and SOC-induced interface splittings are not washed out by alloy disorder. Like $AR_{\rm I}$, the decrease in $\delta$ for Py$|$Pt is more rapid than for Co$|$Pt, $\delta$ going from 0.88 at 100 K to 0.37 at 500 K (a 58\% decrease or $\sim$15\% per 100 K) for Py$|$Pt and from 1.02 at 200 K to 0.61 at 500 K (a 40\% decrease or $\sim$13\% per 100 K) for Co$|$Pt. 

Our results for $AR_{\rm I}$ and $\delta$ show how a combination of alloy, lattice and spin disorder determine how charge and spin currents are transmitted through these interfaces. Transmission is facilitated by increasing bulk disorder. Yet, the combination of alloying, lattice and spin disorder make it difficult to distinguish their individual contributions to the temperature dependence of the interface parameters. In the following subsection, we study the contribution of spin disorder by switching it off for the Py$|$Pt interface.

\subsubsection{Contribution of spin disorder to interface parameters}

\begin{figure}[t]
\includegraphics[width=8.6 cm]{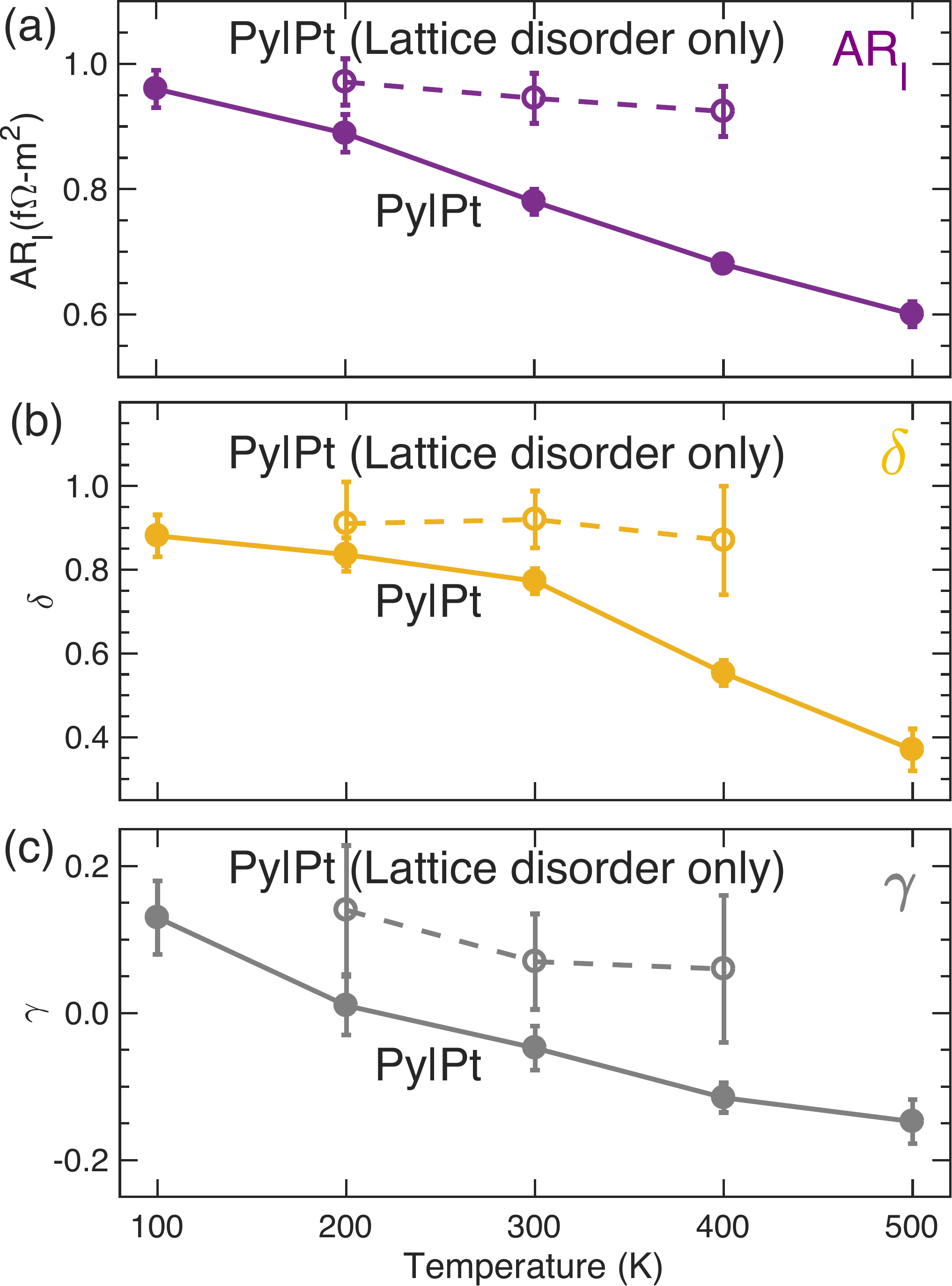}
\caption{Interface parameters $AR_{\rm I},\delta$ and $\gamma$ for Py$|$Pt interface with both lattice and spin disorder (filled circles, solid lines) and Py$|$Pt interface with only lattice disorder (open circles, dashed lines) plotted as a function of temperature.}
\label{figL}
\end{figure}

Including only lattice disorder in Py and Pt at $T=200, 300$ and $400\,$K and keeping the atomic spins in Py ordered at all temperatures, we repeat the calculations for Py$|$Pt. The results for the three interface parameters with only lattice disorder included are compared in \cref{figL} (open circles, dashed lines) with the results that include lattice and spin disorder in Py. We find that the Py$|$Pt interface parameters show a very weak variation with temperature in the absence of any spin disorder, decreasing very slowly with increasing temperature. This weak variation can be attributed to the lattice disorder, but the decrease is much smaller compared to that brought about by spin disorder.

This calculation also highlights that underlying these parameters is the electronic structure mismatch between the two materials making up the interface and the strong spin-orbit coupling in Pt both of which depend only weakly on temperature. This explains why even at 500 K with a significant spin disorder, especially for Py, the interface parameters are not even close to zero. Thus, the common assumption made in interpreting experiments of transparent FM$|$Pt interfaces is not supported by our calculations. We will discuss our results in regard to experiment in \cref{Sec:Exp}.

\subsubsection{Effect of proximity induced magnetization in Pt}

\begin{figure}[t]
\includegraphics[width=8.6 cm]{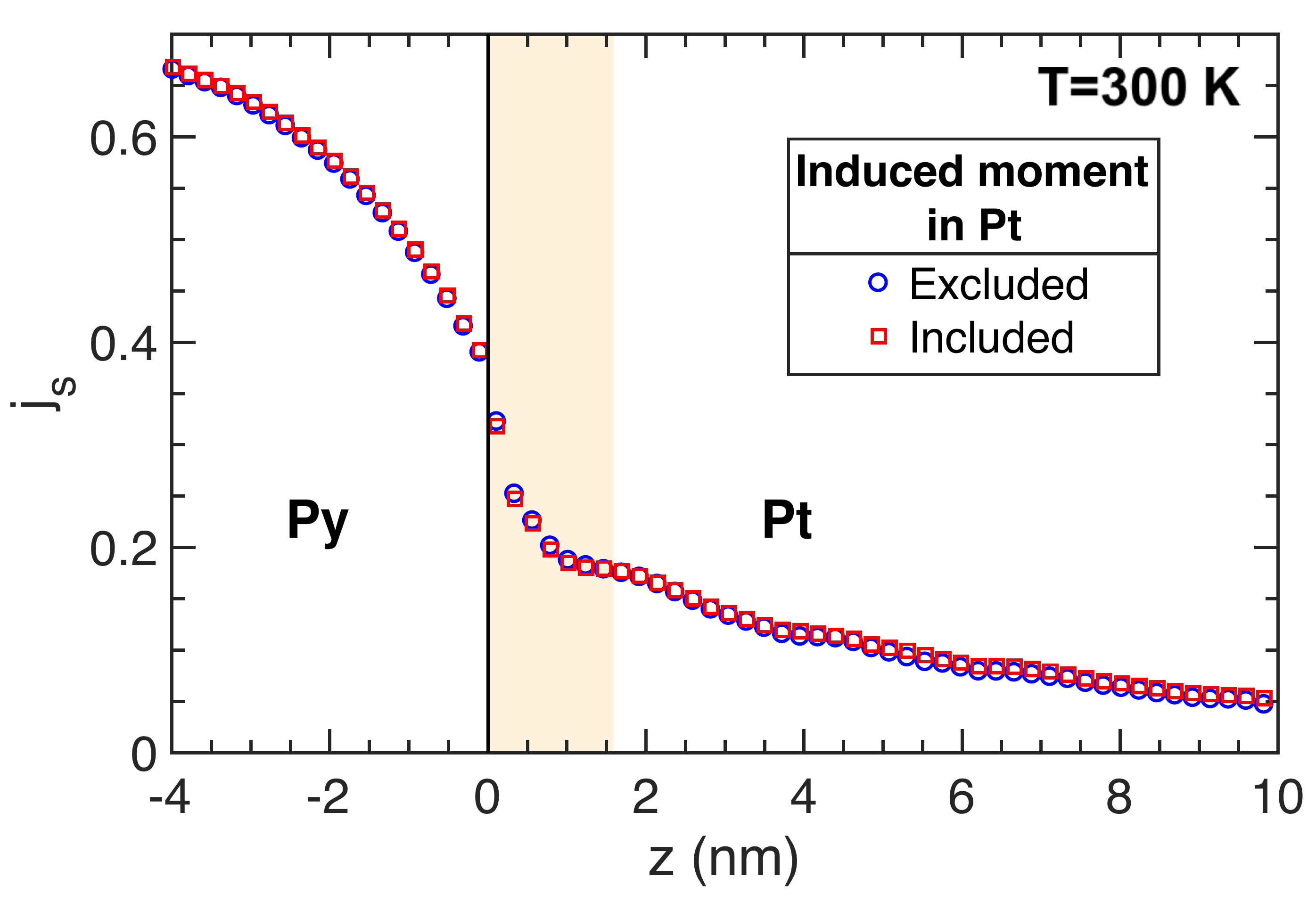}
\includegraphics[width=8.6 cm]{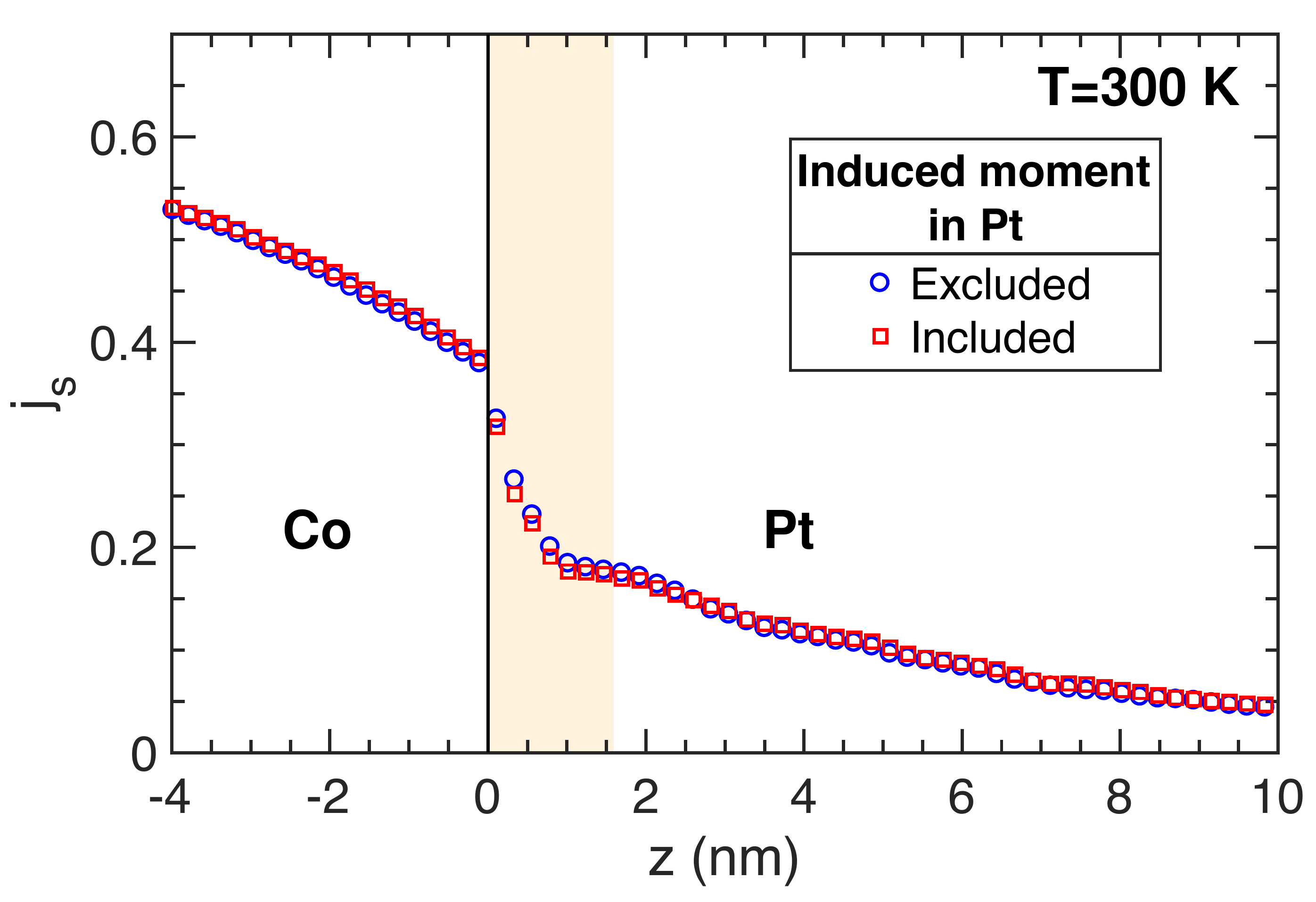}
\caption{Spin currents in (upper panel) Py$|$Pt and (lower panel) Co$|$Pt with (red squares) and without (blue circles) induced magnetic moments in Pt. The yellow shaded region consists of 7 Pt layers adjacent to the FM that that are magnetized. The magnetic Pt potentials are calculated self-consistently for a Pt$|$Co$|$Pt trilayer. The induced moments go from 0.25 $\mu_B$ to 0.0006 $\mu_B$ from left to right in the yellow region.}
\label{figM}
\end{figure}

So far, however, the proximity-induced magnetization of Pt has not been taken into account in our calculations. The proposal that Pt magnetization plays a key role in determining the transport of spins through FM$|$NM interfaces \cite{Huang:prl12, ZhangW:prb15} then poses the question as to how it might affect the interface parameters we have calculated. To address this question, we repeat the RT calculations for both Py$|$Pt and Co$|$Pt interfaces to determine the interface resistance  $AR_{\rm I}$ and the Pt$|$FM$|$Pt  trilayer spin current profile replacing ``bulk, non-magnetic'' Pt potentials adjacent to the FM interface with spin-polarized potentials. As discussed in \cref{sec:Calculations}, these are  obtained from a CPA calculation for a pseudomorphic Pt$|$Co$|$Pt trilayer. Within the errorbar of the calculations, the values of $AR_{\rm I}$ are unchanged. In \cref{figM} we compare the spin current $j_s(z)$ close to the FM$|$Pt interface obtained without (blue circles) and with (red squares) magnetic moments induced in Pt for Py$|$Pt (upper) and Co$|$Pt (lower). We see virtually no change in the spin currents and within the errorbars of the calculation no change in the discontinuity of $j_s(z)$ at the interface; thus $\delta$ and $\gamma$ are not affected by proximity induced magnetism in Pt.     

\section{Comparison with other experiments and calculations}
\label{Sec:Exp}

A direct confrontation of our results for the temperature dependence of the interface parameters with experiment is not possible for a number of reasons.

(1) CPP-MR experiments are conventionally described in terms of the eight VF parameters we have extracted \cite{Bass:jmmm16}. However, the use of superconducting Nb leads means that they are restricted to low temperatures (4.2 K). The multilayers used in these CPP-MR experiments are usually prepared by sputtering and this leads to intermixing of the materials forming the interface rather than atomically sharp interfaces. To extend the present study to intermixed interfaces, the distribution of atomic species about the interface would need to be known but is not. So at present we have no choice but to restrict ourselves to ordered interfaces. 
Alternatively, lateral microstructuring can be used to increase the small resistance of a thin layered structure with respect to long leads \cite{Gijs:prl93}, allowing the interface parameters of epitaxial Co$|$Cu \cite{Oepts:prb96} and sputtered Co$_{50}$Fe$_{50}|$Cu \cite{Delille:jap06} to be determined as a function of temperature up to 300 K. To the best of our knowledge, no similar studies have been carried out for FM$|$Pt interfaces at finite temperatures.

(2) In a spin-pumping plus ISHE (SP-ISHE) experiment, the spins that are driven to precess in the ferromagnet experience enhanced damping at an FM$|$NM interface where the NM material acts as a spin sink. The efficiency of this sink depends on the transparency of the interface as measured by the mixing conductance and spin-dependent interface resistance, on the degree of spin-flipping in the NM bulk as measured by its resistivity and spin-flip diffusion length \cite{Tserkovnyak:prl02a, *Tserkovnyak:prb02b}, and on the interface spin flipping measured by the SML $\delta$ \cite{LiuY:prl14}. The pumped spin current has DC and AC components that are polarized parallel, respectively, perpendicular to the FM magnetization. In most studies, only the DC component of the pumped spin, {\it i.e.} the component parallel to the magnetization, is accessed in terms of the DC inverse spin Hall voltage \cite{Mosendz:prb10, Azevedo:prb11}. Because the polarization of a spin current generated by passing a charge current through a ferromagnetic layer is parallel to the magnetization, the SML we have studied is for spins aligned parallel to the magnetization and our estimates of $\delta$ are in principle suitable for analysing the DC experiments. 
 
However, there is a discrepancy between the modest values of $\delta$ we have found for Py$|$Pt and those found by Liu {\it et al.} who estimated that a value of $\delta=3.7$ was needed to account for the interface damping enhancement \cite{LiuY:prl14} determined from first-principles ``energy pumping'' calculations \cite{Starikov:prb18} that in turn agreed very well with observations \cite{Mizukami:jjap01, *Mizukami:jmmm01}. Because interface spin-orbit coupling may give rise to large interface spin-Hall and inverse spin-Hall effects \cite{WangL:prl16} and affect spins aligned parallel and perpendicular to the magnetization in different ways \cite{Amin:prb16a}, more work is required to understand whether $\delta$ for ``pumped'' spin currents is different from $\delta$ for collinearly spin-polarized currents.

(3) In SHE-STT experiments, an in-plane charge current passing through the NM layer gives rise to a spin current that is polarized perpendicular to the charge current and the interface normal direction. This current would be modified at the FM$|$NM interface by spin memory loss and eventually an interface spin Hall effect. Spin currents are also expected to be generated by spin-orbit filtering and precession \cite{Amin:prl18} at the interface that could exert additional torques on the FM. Such a scenario is described by a phenomenological model \cite{Amin:prb16a, Amin:prb16b} expressed in terms of a set of parameters larger than current experiments are able to evaluate.

It is nevertheless worthwhile briefly discussing experiments that aim to determine bulk parameters such as the spin-flip diffusion length and the spin Hall angle where interface effects may  critically influence the determination of the ``bulk'' parameters. These interface effects are expressed in terms of the parameters we have calculated. For convenience, we have collected the room temperature values in \cref{tab:parameters}.

\begin{table}[t]
\caption{Room temperature transport parameters calculated for Co$|$Pt and Py$|$Pt systems: 
resistivity $\rho$ ($\mu\Omega \,$cm); 
spin-flip diffusion length $l_{\rm sf}$ (nm);
transport polarization $\beta$;
interface resistance $AR_{\rm I} \,({\rm f \,}\Omega {\rm m}^2)$;
spin memory loss $\delta$; interface spin asymmetry $\gamma$. 
}
\begin{ruledtabular}
\begin{tabular}{lldddddd}
    &       &  \multicolumn{3}{c}{Bulk}       & \multicolumn{3}{c} {FM$|$Pt Interface} \\
\cline{3-5}\cline{6-8} 
    &     & \multicolumn{1}{c}{$\rho$} 
          & \multicolumn{1}{c}{$l_{\rm sf}$} 
          & \multicolumn{1}{c}{$\beta$} 
          & \multicolumn{1}{c}{$AR_{\rm I}$}
          & \multicolumn{1}{c}{$\delta$} 
          & \multicolumn{1}{c}{$\gamma$} \\
\cline{1-8}	
Pt & fcc  & 10.8 & 5.3 & 0.0  &      &        &       \\
Co & fcc  &  9.6 & 6.0 & 0.68 & 0.85 & 0.88 & 0.003  \\
Py & fcc  & 15.6 & 2.8 & 0.75 & 0.78 & 0.77 & -0.05  \\
\end{tabular}
\end{ruledtabular}
\label{tab:parameters}
\end{table}

\subsubsection*{Interface resistance and interface spin-asymmetry}

CPP-MR experiments have been used to extract values for $AR_{\rm I}$ and $\gamma$ for the Co$|$Pt interface at 4.2 K by the MSU collaboration of Bass and Pratt \cite{Sharma:jap07, Nguyen:jmmm14}. 
Sharma {\it et al.} \cite{Sharma:jap07} reported $AR_{\rm I}=0.73\pm0.15 {\rm f\,}\Omega {\rm m}^2$ and $\gamma=0.38 \pm 0.06$ but did not include $\delta$ in their VF analysis. 
Nguyen {\it et al.} \cite{Nguyen:jmmm14} reported $AR_{\rm I}=0.53\pm0.20 {\rm f\,}\Omega {\rm m}^2$, $\gamma=0.53 \pm 0.12$ and $\delta_{\rm Co|Pt}=0.9^{+0.5}_{-0.2}$ where the $AR_{\rm I}^*$ parameters have been converted to $AR_{\rm I}$ using the relation $AR_{\rm I}=AR_{\rm I}^*(1-\gamma^2)$. Nguyen's $AR_{\rm I}$ and $\gamma$ values are very different to what we can estimate at low temperature by extrapolation. Their 4.2~K value of $AR_{\rm I}$ is smaller than the RT value we calculate that increases on reducing the temperature and there is no indication that our value of $\gamma$ might shoot up to the large low temperature value that they report.
 
Recently Pham {\it et al.} \cite{Pham:prb21} studied Co$|$Pt, Py$|$Pt (and CoFe$|$Pt) interfaces at room temperature by measuring the spin accumulation generated by the SHE in Pt at two FM electrodes aligned parallel and anti-parallel. They reported the following values: $AR_{\rm I}=13.5\pm2.0 {\rm f\,}\Omega {\rm m}^2$ and $\gamma=0.17\pm0.03$ for Co$|$Pt and $AR_{\rm I}=29.0\pm2.5 {\rm f\,}\Omega {\rm m}^2$ and $\gamma=0.070\pm0.015$ for Py$|$Pt interfaces. In particular, their interface resistances are more than an order of magnitude larger than the values we summarize in \cref{tab:parameters}. When they interpreted their measurements with a model that did not take the interface (resistance, spin asymmetry or spin-flipping) into account, a consistent value of $\Theta_{\rm Pt}l_{\rm Pt}$ (where $\Theta_{\rm Pt} \equiv \Theta_{\rm sH}^{\rm Pt}$) could not be obtained that was independent of the ferromagnetic electrode used. When an interface resistance was included in the model, it was found to be necessary to include an interface spin asymmetry to obtain reasonable values of $\Theta_{\rm Pt}$. The discrepancy between the experimental values and our calculated ones seems to depend on the model used to interpret experiment and on the difficulty evaluating all three interface parameters together with the six bulk parameters (in addition to the bulk spin Hall angles in both materials) in a single sample; if multiple samples are used, there is no guarantee that the interfaces are identical. In the absence of any experimental characterization of the interfaces and correlation with the interface parameters, it is premature to draw any conclusions. 

One further aspect referred to by Pham {\it et al.} that we find troubling is the large value of interface resistance extracted from a computational study of the interface enhancement of the Gilbert damping reported for Py$|$Pt interfaces \cite{LiuY:prl14} that makes use of much of the same computational machinery as used here. We already remarked upon the large value of $\delta$
(and $AR_{\rm I}$) needed to interpret those computational results (that reproduce the experimental damping remarkably well) and speculated on the possibility of there being different $\delta$'s required to describe pumped spin currents ($\delta_\perp$) and collinear spin currents ($\delta_\parallel$). Because the values of interface parameters extracted from experiment depend on using a consistent model containing all relevant parameters, it is clear that more effort, both experimental and theoretical, needs to be devoted to this issue. 



\subsubsection*{Spin memory loss}

Very few experiments/calculations have been carried out at finite temperatures that either take SML into account or recognize its role in determining other bulk parameters. Nguyen {\rm et al.} \cite{Nguyen:jmmm14} carried out CPP-MR measurements for sputtered Co$|$Pt at 4.2 K and reported $\delta_{\rm Co|Pt}=0.9^{+0.5}_{-0.2}$; our RT value of $\delta_{\rm Co|Pt}=0.88$ is seen to increase as the temperature is reduced, \cref{figK}. Earlier, values of $\delta$ in the range 0.2-0.35 were found for Co$|$NM pairs with NM metals whose SOC is weaker than that of Pt \cite{Bass:jpcm07}.

Rojas-Sanchez {\it et al.} \cite{Rojas-Sanchez:prl14} incorporated Nguyen's low temperature interface parameters $\delta$ and $AR_{\rm I}$ for Co$|$Pt into the analysis of their RT SP-ISHE experiments to demonstrate that neglecting interface effects leads to underestimation of bulk parameters. Recently, Tao {\it et al.} \cite{Tao:sca18} reported $\delta$ for Py$|$Pt and Co$|$Pt as 0.63$\pm$0.05 and 0.39$\pm$0.01 respectively from SP-ISHE experiments. Berger {\it et al.} \cite{Berger:prb18b} attributed a $\sim$60\% loss of damping enhancement measured using a ``Vector Network Analyzer-FMR'' technique at a Py$|$Pt interface to SML. We note that the  resistivity and inverse SDL measured for these samples far exceed what can be attributed to  electron-phonon scattering indicating that this may not be the dominant scattering mechanism \cite{Wesselink:prb19} making a direct comparison of our calculated value of $\delta$ with the value extracted from experiment problematic. 
Zhang {\it et al.} \cite{ZhangW:natp15} introduced an interface transparency parameter to measure the efficiency of spin-Hall induced spin current transfer from the NM metal to the FM metal. From ST-FMR measurements, they predict a smaller interface transparency of 0.25$\pm$0.05 for Py$|$Pt than 0.65$\pm$0.06 for Co$|$Pt which is the opposite of the trend we found for the interface resistance. It should be noted that the transparency defined by Zhang {\it et al.} as a measure of the spin-Hall torque efficiency involves many more factors than just the SML e.g., the mixing conductance, the interfacial contribution to the spin Hall effect, etc.

Such or similar spin transparency parameters have been introduced by a number of workers  \cite{Rojas-Sanchez:prl14, ZhangW:natp15, Nguyen:prl16, Zhu:prap20, Belashchenko:prl16, Dolui:prb17} but are not uniquely defined. Dolui states a relation between the phenomenological parameter $\zeta$ introduced in \cite{Rojas-Sanchez:prl14} and $\delta$ but evaluating it would  require extracting additional system parameters from their calculations. The values that Belashchenko finds for $\delta$ for Cu$|$Pd (0.38-0.54) are in reasonable agreement with experiment (0.24), albeit overestimated. Both sets of calculations consider an interface between ballistic leads, assuming an equilibrium distribution for the incident electrons in their scattering calculation. This assumption is questionable since generally there is a non-equilibrium distribution close to the interface. Belashchenko takes this effect partly into account by using a renormalization introduced in \cite{Schep:prb97}, but the non-equilibrium distribution is (i) in the diffusive limit affected by disorder not taken into account in those calculations, and (ii) in the ballistic limit it is affected by other interfaces at distances comparable to the mean free path $\lambda$. The present work is in the diffusive limit (distance between interfaces $>> \lambda$) and does take into account the effect of non-equilibrium distributions close to the interfaces consistent with the bulk disorder. The ansatz introduced by Schep {\it et al.} does not readily lend itself to including a temperature dependence.

\section{Conclusions}

We have developed a practical scheme for calculating the temperature dependence of bulk and interface transport parameters for real materials incorporating all of the complexity of the electronic structure of transition metals both nonmagnetic and magnetic. We illustrated it with a study of Py$|$Pt and Co$|$Pt interfaces for which we determined all eight parameters contained in the generalized semiclassical VF model used to interpret experiment. The remarkably good fit of the spin currents calculated from first-principles scattering theory with the VF model give us every reason to believe in the meaningfulness of the parameter values we extract. Bulk spin-flip diffusion lengths and transport polarizations could be determined for Py, Co and Pt independent of interface contributions. Thermal lattice and spin disorder facilitates the transmission of electrons through the interface, yet does not render $AR_{\rm I}$ and $\delta$ zero even at high temperatures. Finally, the magnetization induced in Pt by proximity to a ferromagnet does not affect the interface discontinuity.  

{\it Acknowledgements.---}This work was financially supported by the ``Nederlandse Organisatie voor Wetenschappelijk Onderzoek'' (NWO) through the research programme of the former ``Stichting voor Fundamenteel Onderzoek der Materie,'' (NWO-I, formerly FOM) and through the use of supercomputer facilities of NWO ``Exacte Wetenschappen'' (Physical Sciences). K.G. acknowledges funding from the Shell-NWO/FOM “Computational Sciences for Energy Research” PhD program (CSER-PhD; nr.~i32; project number 13CSER059) and is grateful to Yi Liu for help in starting this work and to S. Wildeman for helpful discussions. The work was also supported by the Royal Netherlands Academy of Arts and Sciences (KNAW). Work in Beijing was supported by the National Natural Science Foundation of China (Grant No. 61774018), the Recruitment Program of Global Youth Experts, and the Fundamental Research Funds for the Central Universities (Grant No. 2018EYT03).


%

\end{document}